# Terahertz parametric amplification as a reporter of exciton condensate dynamics

Sheikh Rubaiat Ul Haque[1], Marios H. Michael[2], Junbo Zhu[3], Yuan Zhang[1], Lukas Windgätter[4], Simone Latini[4], Joshua P. Wakefield[3], Gu-Feng Zhang[1], Jingdi Zhang[1,5], Angel Rubio[4,6,7], Joseph G. Checkelsky[3], Eugene Demler[2,8], Richard D. Averitt[1*]

**Affiliations**

[1]Department of Physics, University of California San Diego, La Jolla, CA 92093, USA.

[2]Department of Physics, Harvard University, Cambridge, MA 02138, USA.

[3]Department of Physics, Massachusetts Institute of Technology, Cambridge, MA 02139, USA.

[4]Max Planck Institute for the Structure and Dynamics of Matter (MPSD), Luruper Chausse 149, 22761 Hamburg, Germany.

[5]Department of Physics, The Hong Kong University of Science and Technology, Clear Water Bay, Kowloon, Hong Kong, China.

[6]Center for Computational Quantum Physics, The Flatiron Institute, 162 Fifth Avenue, New York, NY 10010, USA.

[7]Nano-Bio Spectroscopy Group, Departamento de Física de Materiales, Universidad del País Vasco, 20018 San Sebastían, Spain.

[8]Institute for Theoretical Physics, ETH Zürich, 8093 Zürich, Switzerland.

*e-mail: raveritt@ucsd.edu

**Abstract** Condensates are a hallmark of emergence in quantum materials with superconductors and charge density wave as prominent examples. An excitonic insulator (EI) is an intriguing addition to this library, exhibiting spontaneous condensation of electron-hole pairs[1]. However, condensate observables can be obscured through parasitic coupling to the lattice[2-6]. Time-resolved terahertz (THz) spectroscopy can disentangle such obscurants through measurement of the quantum dynamics. We target $Ta_2NiSe_5$, a putative room-temperature EI where electron-lattice coupling dominates the structural transition ($T_c = 326$ K), hindering identification of excitonic correlations[6-15]. A pronounced increase in the THz reflectivity manifests following photoexcitation and exhibits a BEC-like temperature dependence. This occurs well below the $T_c$, suggesting a novel approach to monitor exciton condensate dynamics. Nonetheless, dynamic condensate-phonon coupling remains as evidenced by peaks in the enhanced reflectivity spectrum at select infrared-active phonon frequencies. This indicates that parametric reflectivity enhancement arises from phonon squeezing, validated using Fresnel-Floquet theory and density functional calculations. Our results highlight that coherent dynamics can drive parametric stimulated emission with concomitant possibilities, including entangled THz photon generation.

In small bandgap semiconductors or semimetals, coulomb interactions between electrons and holes can result in pairing correlations, leading to ground-state exciton condensation that exhibits macroscopic phase coherence with an associated interaction-induced insulating gap[1]. While exciton condensates have been observed in quantum Hall bilayers[16] and semiconductor quantum wells[17], a pure excitonic insulator (EI) of intrinsic origin remains elusive. Electron-lattice interactions play a primary role both in disrupting the purity of an EI state and in hiding observable consequences of exciton condensation. For example, charge density modulation can arise from exciton condensation as recently shown in indirect-gap transition metal dichalcogenides[18,19]. It is imperative, therefore, to delineate the full ramifications of electron-phonon coupling which includes dynamical attributes[20].

The layered quasi-1D direct-gap semiconductor $Ta_2NiSe_5$ (TNS) has drawn considerable interest as a potential EI material[2-13]. Excitons in TNS arise from charge transfer between parallel chains of Ta and Ni atoms along the $a$-axis (Fig. 1a). A second order monoclinic-orthorhombic phase transition occurs at $T_c = 326$ K, with the formation of a direct gap (~160 meV, Fig. 1b) that increases in magnitude as the temperature is lowered[2-10]. A plethora of experiments have been performed to clarify the structural or electronic dominance of the phase transition, albeit with contradictory results and interpretations[2,4,5,10-15,21-34]. The accumulated evidence suggests that TNS might harbor a room temperature EI phase which is driven by a structural instability[7,10,12,13,21,22,27,29,31-35].

A full accounting of interactions in TNS necessitates experiments capable of monitoring condensate-lattice dynamics[20,33] and progress investigating TNS utilizing various time-resolved optical techniques has been made[3,7,9,21]. In this report, we utilize time-resolved terahertz (THz) spectroscopy as this technique is uniquely suited to probe condensate dynamics with discriminatory power arising from the temporal response coupled with broadband spectral access of the low energy electrodynamics. As Higgs mode and pair density wave studies of superconductors have shown, THz experiments can access nonlinear optical phenomena that do not manifest in the linear response[36,37]. We demonstrate that time-resolved THz spectroscopy reports unambiguous condensate signatures while simultaneously clarifying the dynamical repercussions of electron-phonon coupling in TNS.

We performed time-resolved reflection-based broadband THz spectroscopy on TNS single crystals, measuring the dynamics following above-gap photoexcitation (Fig. 1b) with 0.5 eV infrared pump pulses with 45 fs pulse width (Extended Data Fig. 2). Previous studies[3,7,21] indicate a stronger condensate response parallel to the atomic chains (along the $a$-axis), irrespective of pump polarization. For our measurements, the probe polarization is along the $a$-axis, while the pump is polarized along the $c$-axis. The black line in Fig. 1c shows the equilibrium $a$-axis reflectivity of TNS at 90 K, revealing multiple phonons in the spectrum (positions indicated by vertical gray dashed lines) in accordance with infrared measurements[8,30]. Figures 1c – e show the corresponding pump-induced relative reflectivity enhancement $(R + \Delta R)/R$ at three different temperatures, 1 ps after photoexcitation. A broadband fluence dependent signal is evident with a magnitude that decreases with increasing temperature, nearly vanishing at 270K. Importantly, the reflectivity enhancement peaks in the vicinity of the phonon resonances. For example, at 90K (Fig. 1c), an increase of approximately 25% is evident at 4.7 THz (corresponding to a $B_{3u}$ phonon) that, as described below, plays a prominent role in the dynamics through preferential coupling to the band structure in the monoclinic phase. There is also a 9% increase centered at 1.8 THz (orange arrows) which does not overlap with any phonons.

Figures 2a and 2b plots the temperature dependence of the peak pump-induced reflectivity change $\Delta R/R$ at frequencies 1.8 THz and 4.7 THz, revealing a clear roll-off with increasing temperature. Focusing on the 4.7 THz data, the temperature dependence of $\Delta R/R$ is fit using $\Delta R/R(T) = \frac{A/\Delta_E}{1+B\,exp\,(-\frac{\Delta_E}{k_B T})}$ (dashed curves in Fig. 2b), where A and B are parameters and $\Delta_E$ is a temperature-independent energy scale[22,38,39]. The fits – shown in Fig. 2b as dashed lines – yield $\Delta_E = 113 \pm 7$ meV ($\Delta_E = 120 \pm 13$ meV) for a fluence of 0.2 mJ/cm$^2$ (0.4 mJ/cm$^2$), in reasonable agreement with the gap (~160 meV) from optical conductivity and ARPES studies[7,10]. Moreover, specific heat analysis reveals that the pump-induced temperature rise is small (<5K for 0.4 mJ/cm$^2$), ruling out a thermal origin of the signal (Extended Data Fig. 6a – b). The absolute reflectivity increase $\Delta R$ can be obtained by multiplying $\Delta R/R$ by the equilibrium reflectivity as shown in Fig. 2c. $\Delta R$ is proportional to the increase in the number of reflected THz photons. The integral of $\Delta R$ over the spectral range 0.5 – 7.5 THz (shaded areas in Fig. 2c; $\Delta \eta = \frac{1}{2\pi} \int_{0.5\,\text{THz}}^{7.5\,\text{THz}} \Delta R(\omega)\,d\omega$) is plotted in Fig. 2d and exhibits similar temperature behavior as $\Delta R/R$. Fitting $\Delta \eta(T)$ with the same equation we used for $\Delta R/R(T)$, we obtain, $\Delta_E = 115 \pm 14$ meV.

Collectively, the data in Fig. 2 suggests photoinduced perturbation of a BEC-like condensate response which results in broadband electronic dynamics encoded in the enhanced THz reflectivity, occurring well below the structural transition temperature. The spectra also reveal dynamic condensate-phonon coupling as indicated by peaks in the enhanced reflectivity spectrum at select infrared-active phonon frequencies. Furthermore, that the $\Delta R/R$ temperature behavior resembles that of an underdoped cuprate with a pseudogap where preformed pairs exist above $T_C$ but condensation takes place only below $T_C$[38]. The gap represents the pair binding energy and is temperature-independent, a key ingredient for a BEC-like phase transition. Similarly, in terms of exciton picture in TNS, the notion of preformed excitons and BEC-like phase transition is consistent with previous near-infrared and ARPES studies[4,10,21,27,29,31,35].

The data showing the temporal response of the photoinduced condensate perturbation are presented in Fig. 3. Fig. 3a – b display one-dimensional pump-probe scans of the pump-induced change in the peak electric field of the THz pulse as a function of pump-probe delay time for different temperatures and fluences. The recovery dynamics exhibit a short-lived response of ~2 ps and were fitted with a single-exponential decay. The temperature dependence of the recovery time $\tau$ from the fits for 0.2 mJ/cm$^2$ and 0.4 mJ/cm$^2$ fluences are plotted in Fig. 3c and 3d. With decreasing temperature $\tau$ increases, akin to dynamics associated with a gap as observed in superconductors and related materials[22,38]. The dashed lines in Fig. 3c, 3d are fits to $\tau(T)$ utilizing a phenomenological temperature independent BEC gap model[38,39] yielding $\Delta_E = 110 \pm 5$ meV, consistent with $\Delta_E$ as obtained from the $\Delta R/R$ data. Deviation from the model at higher temperatures can be attributed to opening of additional decay channels due to phonon scattering or BCS-like partial closing of the gap[22,29]. The dynamics obtained from the spectral response (i.e., full two-dimensional scans) are consistent with the 1-d scans. Fig. 3e and 3g display $\Delta R/R(T)$ and $\Delta\eta(T)$ at different pump-probe delays for two different temperatures (90 K and 180 K) at 0.2 mJ/cm$^2$ fluence, while Fig. 3f and 3h plot the 0.4 mJ/cm$^2$ fluence results. All data have been rescaled and plotted on the same figure with the left and right panels representing $\Delta R/R(T)$ (at 4.7 THz, solid squares) and $\Delta\eta(T)$ (open circles), respectively. The dashed black lines correspond to the single-exponential decay fits at respective temperatures and fluences obtained from Fig. 3a, 3b. We see that both $\Delta R/R(T)$ and $\Delta\eta(T)$ closely match the fits. Overall, the fast dynamics of $\Delta R/R(T)$ and $\Delta\eta(T)$ further confirms the electronic origin of the reflectivity enhancement.

Insight into the experimentally observed reflectivity dynamics can be obtained from a minimal phenomenological model based on Fresnel-Floquet theory[40]. Briefly, the sample permittivity $\varepsilon$ acquires an oscillatory component upon photoexcitation given as $\delta\varepsilon(t) \propto \cos(\omega_d t)$, where $\omega_d$ is the driving frequency (importantly, $\omega_d$ is distinct from the pump photon energy). Plugging the dynamic permittivity (i.e., $\varepsilon = \varepsilon_{eq} + \delta\varepsilon(t)$, where $\varepsilon_{eq}$ is the equilibrium permittivity) into the dispersion relation determined from Maxwell's equations and applying the appropriate boundary conditions (see Eqns. 1 – 9 in Methods) enables calculation of $\Delta R/R$. To fit the experimental $\Delta R/R$ data, $\varepsilon_{eq}$ is obtained from the experimentally measured reflectivity and taking $\omega_d$ = 9.4 THz. This is twice the frequency of the 4.7 THz ($\omega$) $B_u$ phonon and is fully justified using an effective Hamiltonian description and DFT calculations below. The fit of $\Delta R/R$ is shown in Fig. 4a and is in good agreement with experiment, highlighting that the dominant contribution to $\omega_d$ is the 4.7 THz phonon. The deviation from experiment likely arises from neglecting other IR phonons that participate (to a lesser degree) in the dynamics. We note that no coherent oscillation was observed in the transient electric field response as depicted in Fig. 3a – b. The reason is that although the pump pulses have the necessary bandwidth (~ 9.8 THz) to drive the 9.4 THz oscillations, the THz probe does not have enough bandwidth to be able to fully resolve the fast oscillations.

Our analysis reveals that the dynamically evolving material properties act as a parametric drive at $\omega_d = 2\omega$ (depicted in Fig. 4b). Specifically, an incident THz probe photon at frequency $\omega_s$ interacts with the dynamic interface to generate reflected photons at the signal ($\omega_s$) and idler frequencies ($\omega_{id} = \omega_d - \omega_s$). Thus, $\Delta R/R$ is effectively the stimulated parametric emission spectrum. Importantly, our results show that Fresnel-Floquet analysis provides an approach to describe coherent nonlinear effects, in contrast to analysis of the dynamics as "frozen snapshots" of the optical conductivity at each pump-probe delay. Thermal shift of phonons and phonon screening by photoelectrons were considered as an alternative explanation for the reflectivity enhancement. However, shifting of phonon frequency leads to both negative and positive changes in reflectivity (See Methods and Extended Data Fig. 7). In contrast, an overall reflectivity increase in THz regime was observed which was symmetrically centered around 4.7 THz, indicating that parametric drive at twice the IR phonon frequency is the most natural mechanism that can support the experimental observation.

While Fresnel-Floquet theory provides a good fit to the data, further analysis is required to justify $\omega_d = 9.4$ THz since IR-active phonons are not directly excited by the pump. Electron-phonon coupling provides a mechanism whereby photoexcitation with above-gap 0.5 eV pulses polarized along the *a*-axis can couple to the IR modes. This is captured by an effective electron-phonon Hamiltonian $H_{e-ph,eff} = \sum_k g_k Q_{IR}^2 (n_{ck} - n_{dk})$ where $Q_{IR}$ is the phonon coordinate, $(n_{ck} - n_{dk})$ represents the photoexcited electron occupation and $g_k$ is the effective electron-phonon coupling. Since IR phonons are not directly excited by the off-resonant pump, the expectation value of the phonon mode is zero (i.e., $\langle Q_{IR} \rangle = 0$). However, above-gap photoexcitation (see Eqns. 10 – 14 in Methods) leads to squeezed phonons that coherently oscillate at twice the phonon frequency $2\omega$, $\langle Q_{IR}^2(t) \rangle = \langle Q_{IR}^2 \rangle_0 + A \cos(2\omega t)$, as shown in Fig. 4c. Although, this $2\omega$ overtone does not have a dipole due to its symmetry and thus, cannot emit THz radiation, it can create stimulated emission (Extended Data Fig. 8). In short, the $2\omega$ phonon squeezing oscillation serves as the driving frequency $\omega_d$, creating a photon pair with opposite momenta whose frequencies satisfy the parametric resonance relation, $\omega_s(k) + \omega_{id}(-k) = \omega_d = 2\omega$. To sum up, above-gap photoexcitation initiates phonon dynamics that result in enhanced THz reflectivity.

In principle, any of the IR phonons can couple to the electrons to enable parametric enhancement of the reflectivity as discussed in the previous paragraph. To investigate this, DFT calculations were performed in the frozen phonon limit where the band structure is calculated for lattice displacements corresponding to various phonon eigenmodes. Further details and discussion are provided in Methods which includes Extended Data Fig. 10a – b and Table 1 for clarity. The DFT results are pictorially shown in Fig. 4d and 4e in the vicinity of Γ point. The calculations reveal that for the 4.7 THz $B_u$ phonon, electron-phonon coupling in the low-temperature monoclinic phase is the strongest (Extended Data Fig. 9) with small displacements of the phonon coordinate $Q_{IR}$ leading to a band shift as in Fig. 4d and Extended Data Fig. 10a. The other phonons in the vicinity of 4.7 THz do not couple nearly as strongly. Moreover, the calculations reveal no significant phonon-induced band shift in the high-temperature orthorhombic phase (Fig. 4e, Extended Data Fig. 10b). Thus, DFT calculations provide a microscopic basis for the experimentally observed THz stimulated emission spectrum in Fig. 2c and 4a described using phonon squeezing and Fresnel-Floquet theory. Furthermore, DFT reveals that the electron-phonon coupling is very sensitive to the phase transition in TNS.

This framework, in conjunction with the temperature dependence of the parametric reflectivity enhancement (Fig. 1, 2), reveals that the phonons are strongly coupled with the exciton condensate. In particular, the experimental temperature dependent results are consistent with $g_k \propto \Delta^2_{TNS}$. That is, the electron-phonon coupling $g_k$ is proportional to the square of the order parameter $\Delta_{TNS}$ as embodied in the interaction induced gap. This may involve a contribution from a structural order parameter intertwined with the excitonic order parameter[10,20,24], though we note that the observed temperature dependence of the enhanced reflectivity occurs well below $T_c$. Thus, while the monoclinic phase promotes strong electron-phonon coupling, we expect the dominant contribution to originate from the exciton condensate.

These results, combined with previous reports about formation of an exciton condensate at monoclinic phase[3-7,10,13,21,26,27,29,32,35], demonstrate that nonlinear THz parametric stimulated emission serves as a sensitive reporter of condensate dynamics and the coupling to dynamically fluctuating degrees of freedom. Our results also unveil the first spectral fingerprints of an exciton condensate in terms of an IR phonon response. Notably, phonon squeezing provides a route to drive the condensate without direct excitation of the Higgs mode. In addition, the momentum dependence of $g_k$ motivates future experiments to measure the reflectivity enhancement as a function of pump photon energy. Our experimental technique and the application of Fresnel-Floquet theory are applicable to a host of other quantum materials to gain new insights into subtle interactions that modify the condensate response from textbook expectations. Moreover, condensate-driven THz parametric amplification may lead to novel photon sources and optoelectronic devices using EIs. Further, our results suggest intriguing potential for novel single-photon THz quantum optics, paving the way towards EI-based THz lasers that can drive innovation in quantum information and sensing technology.

**References and Notes**

1. Jérome, D., Rice, T. & Kohn, W., Excitonic insulators. *Phys. Rev.* **158**, 462 (1967). doi:10.1103/PhysRev.158.462

2. Kaneko, T., Toriyama, T., Konishi, T. & Ohta, Y., Orthorhombic-to-monoclinic phase transition induced by the Bose-Einstein condensate of excitons. *Phys. Rev. B* **87**, 035121 (2013). doi:10.1103/PhysRevB.87.035121

3. Werdehausen, D., Takayama, T., Höppner, M., Albrecht, G., Rost, A.W., Lu, Y.F., Manske, D., Takagi, H. & Kaiser, S., Coherent order parameter oscillations in the ground state of the excitonic insulator $Ta_2NiSe_5$. *Sci. Adv.* **4**, eaap8652 (2018). doi:10.1126/sciadv.aap8652

4. Sugimoto, K., Nishimoto, S., Kaneko, T. & Ohta, Y., Strong coupling nature of the excitonic insulator state in $Ta_2NiSe_5$. *Phys. Rev. Lett.* **120**, 247602 (2018). doi:10.1103/PhysRevLett.120.247602

5. Mazza, G., Rösner, M., Windgätter, L., Latini, S., Hübener, H., Millis, A. J., Rubio, A. & Georges, A., Nature of symmetry breaking at the excitonic insulator transition: $Ta_2NiSe_5$. *Phys. Rev. Lett.* **124**, 197601 (2020). doi:10.1103/PhysRevLett.124.197601

6. Lu, Y.F., Kono, H., Larkin, T. I., Rost, A., Takayama, T., Boris, A.V., Keimer, B. & Takagi, H., Zero-gap semiconductor to excitonic insulator transition in $Ta_2NiSe_5$. *Nat. Commun.* **8**, 14408 (2017). doi:10.1038/ncomms14408

7. Larkin, T. I., Yaresko, A. N., Pröpper, D., Kikoin, K. A., Lu, Y.F., Takayama, T., Mathis, Y. L., Rost, A. W., Takagi, H., Keimer, B. & Boris, A. V., Giant exciton Fano resonance in quasi-one-dimensional $Ta_2NiSe_5$. *Phys. Rev. B* **95**, 195144 (2017). doi:10.1103/PhysRevB.95.195144

8. Larkin, T. I., Dawson, R. D., Höppner, M., Takayama, T., Mathis, Y. L., Takagi, H., Keimer, B. & Boris, A. V., Infrared phonon spectra of quasi-one-dimensional $Ta_2NiSe_5$ and $Ta_2NiS_5$. *Phys. Rev. B* **98**, 125113 (2018). Doi:10.1103/PhysRevB.98.125113

9. Mor, S., Herzog, M., Noack, J., Katayama, N., Nohara, M., Takagi, H., Trunschke, A., Mizokawa, T., Monney, C. & Stähler, J., Inhibition of the photoinduced structural phase transition in the excitonic insulator $Ta_2NiSe_5$. *Phys. Rev. B* **97**, 115154 (2018). doi:10.1103/PhysRevB.97.115154

10. Seki, K., Wakisaka, Y., Kaneko, T., Toriyama, T., Konishi, T., Sudayama, T., Saini, N. L., Arita, M., Namatame, H., Taniguchi, M., Katayama, N., Nohara, M., Takagi, H., Mizokawa, T. & Ohta, Y., Excitonic Bose-Einstein Condensation in $Ta_2NiSe_5$ above room temperature. *Phys. Rev. B* **90**, 155116 (2014). doi:10.1103/PhysRevB.90.155116

11. Nakano, A., Hasegawa, T., Tamura, S., Katayama, N., Tsutsui, S. & Sawa, H., Antiferroelectric distortion with anomalous phonon softening in the excitonic insulator $Ta_2NiSe_5$. *Phys. Rev. B* **98**, 045139 (2018). doi:10.1103/PhysRevB.98.045139

12. Saha, T., Golež, D., Ninno, G.D., Mravlje, J., Murakami, Y., Ressel, B., Stupar, M. & Ribič, P. R., Photoinduced phase transition and associated timescales in the excitonic insulator $Ta_2NiSe_5$. *Phys. Rev. B* **103**, 144304 (2021). doi:10.1103/PhysRevB.103.144304

13. Lee, J., Kang, C.J., Eom, M. J., Kim, J. S., Min, B. I. & Yeom, H. W., Strong interband interaction in the excitonic insulator phase of $Ta_2NiSe_5$. *Phys. Rev. B* **99**, 075408 (2019). doi:10.1103/PhysRevB.99.075408
12

14. Kim, K., Kim, H., Kim, J., Kwon, C., Kim, J. S. & Kim, B. J., Direct observation of excitonic instability in $Ta_2NiSe_5$. *Nat. Commun.* **12**, 1969 (2021). doi:10.1038/s41467-021-22133-z

15. Kim, M.J., Schultz, A., Takayama, T., Isobe, M., Takagi, H. & Kaiser, S., Phononic soft mode behavior and a strong electronic background across the structural phase transition in the excitonic insulator $Ta_2NiSe_5$. *Phys. Rev. Research* **2**, 042039I (2020). doi:10.1103/PhysRevResearch.2.042039

16. Eisenstein, J. P., Exciton condensation in bilayer quantum Hall systems. *Annu. Rev. Condens. Matter Phys.* **5**, 159-181 (2014). doi:10.1146/annurev-conmatphys-031113-133832

17. Amo, A., Lefrère, J., Pigeon, S., Adrados, C., Ciuti, C., Carusotto, I., Houdré, R., Giacobino, E. & Bramati, A., Superfluidity of polaritons in semiconductor microcavities. *Nat. Phys.* **5**, 805-810 (2009). doi:10.1038/nphys1364

18. Kogar, A., Rak, M. S., Vig, S., Husain, A. A., Flicker, F., Joe, Y. I., Venema, L., MacDougall, G. J., Chiang, T. C., Fradkin, E., Wenzel, J. v. & Abbamonte, P., Signatures of exciton condensation in a transition metal dichalcogenide. *Science* **358**, 1314-1317 (2017). doi:10.1126/science.aam6432

19. Song, Y., Jia, C., Xiong, H., Wang, B., Jiang, H., Huang, K., Hwang, J., Li, Z., Hwang, C., Liu, Z., Shen, D., Sobota, J. A., Kirchmann, P., Xue, J., Devereaux, T. P., Mo, S.K., Shen, Z.X. & Tang, S., Evidences for the exciton gas phase and its condensation in monolayer 1T-$ZrTe_2$. *Nat. Commun.* **14**, 1116 (2023). doi:10.1038/s41467-023-36857-7

20. Golež, D., Sun, Z., Murakami, Y., Georges & A., Millis, A. J., Nonlinear spectroscopy of collective modes in an excitonic insulator. *Phys. Rev. Lett.* **125**, 257601 (2020). doi:10.1103/PhysRevLett.125.257601

21. Bretscher, H. M., Andrich, P., Telang, P., Singh, A., Harnagea, L., Sood, A. K. & Rao, A., Ultrafast melting and recovery of collective order in the excitonic insulator $Ta_2NiSe_5$. *Nat. Commun.* **12**, 1699 (2021). doi:10.1038/s41467-021-21929-3

22. Werdehausen, D., Takayama, T., Albrecht, G., Lu, Y.F., Takagi, H. & Kaiser, S., Photo-excited dynamics in the excitonic insulator $Ta_2NiSe_5$. *J. Phys.: Condens. Matter* **30**, 305602 (2018). doi:10.1088/1361-648X/aacd76

23. Baldini, E., Zong, A., Choi, D., Lee, C., Michael, M. H., Windgätter, L., Mazin, I. I., Latini, S., Azoury, D., Lv, B., Kogar, A., Wang, Y. Lu, Y., Takayama, T., Takagi, H., Millis, A. J., Rubio, A., Demler, E. & Gedik, N., The spontaneous symmetry breaking in $Ta_2NiSe_5$ is structural in nature. *Preprint at* https://arxiv.org/abs/2007.02909 (2020).

24. Subedi, A., Orthorhombic-to-monoclinic transition in $Ta_2NiSe_5$ due to a zone-center optical phonon instability. *Phys. Rev. Materials* **4**, 083601 (2020). doi:10.1103/PhysRevMaterials.4.083601

25. Windgätter, L., Rösner, M., Mazza, G., Hübner, H., Georges, A., Millis, A. J., Latini, S. & Rubio, A., Common microscopic origin of the phase transition in $Ta_2NiS_5$ and the excitonic insulator candidate $Ta_2NiSe_5$. *Npj Comput. Mater.* **7**, 210 (2021). doi:10.1038/s41524-021-00675-6

26. Volkov, P. A., Ye, M., Lohani, H., Feldman, I., Kanigel, A. & Blumberg, G., Critical charge fluctuations and quantum coherent state in excitonic insulator $Ta_2NiSe_5$. *Npj Quantum Mater*. **6**, 52 (2021). doi:10.1038/s41535-021-00351-4

27. Pal, S., Grover, S., Harnagea, L., Telang, P., Singh, A., Muthu, D. V. S., Waghmare, U. V. & Sood, A.K., Destabilizing excitonic insulator phase by pressure tuning of exciton phonon coupling. *Phys. Rev. Research* **2**, 043182 (2020). doi:10.1103/PhysRevResearch.2.043182

28. Tang, T., Wang, H., Duan, S., Yang, Y., Huang, C., Guo, Y., Qian, D. & Zhang, W., Non-Coulomb strong electron-hole binding in $Ta_2NiSe_5$ revealed by time- and angle-resolved photoemission spectroscopy. *Phys. Rev. B* **101**, 235148 (2020). doi:10.1103/PhysRevB.101.235148

29. Chen, L., Han, T. T., Cai, C., Wang, Z. G., Wang, Y. D., Xin, Z. M. & Zhang, Y., Doping-controlled transition from excitonic insulator to semimetal in $Ta_2NiSe_5$. *Phys. Rev. B* **102**, 161116I (2020). doi:10.1103/PhysRevB.102.161116

30. Seo, Y.S., Eom, M. J., Kim, J. S., Kang, C.J., Min, B.I. & Hwang, J., Temperature-dependent excitonic superfluid plasma frequency evolution in an excitonic insulator. *Sci. Rep.* **8**, 11961 (2018). doi:10.1038/s41598-018-30430-9

31. Wakisaka, Y., Sudayama, T., Takubo, K., Mizokawa, T., Arita, M., Namatame, H., Taniguchi, M., Katayama, N., Nohara, M. & Takagi, H., Excitonic insulator state in $Ta_2NiSe_5$ probed by photoemission spectroscopy. *Phys. Rev. Lett.* **103**, 026402 (2009). doi:10.1103/PhysRevLett.103.026402

32. Bretscher, H. M., Andrich, P., Murakami, Y., Golež, D., Remez, B., Telang, P., Sing, A., Harnagea, L., Cooper, N. R., Millis, A. J., Werner, P., Sood & A. K., Rao, A., Imaging the coherent propagation of collective modes in the excitonic insulator $Ta_2NiSe_5$ at room temperature. *Sci. Adv.* **7**, eabd6147 (2021). doi:10.1126/sciadv.abd6147

33. Golež, D., Dufresne, S. K. Y., Kim, M. J., Boschini, F., Chu, H., Murakami, Y., Levy, G., Mills, A. K., Zhdanovich, S., Isobe, M., Takagi, H., Kaiser, S., Werner, P., Jones, D. J., Georges, A., Damascelli, A. & Millis, A. J., Unveiling the underlying interactions in $Ta_2NiSe_5$ from photo-induced lifetime change. *Phys. Rev. B* **106**, L121106 (2022). doi:10.1103/PhysRevB.106.L121106

34. Ye, M., Volkov, P. A., Lohani, H., Feldman, I., Kim, M., Kanigel, A. & Blumberg, G., Lattice dynamics of the excitonic insulator $Ta_2Ni(Se_{1-x}S_x)_5$. *Phys. Rev. B* **104**, (045102 (2021). doi:10.1103/PhysRevB.104.054102

35. Fukutani, K., Stania, R., Kwon, C. I., Kim, J. S., Kong, K. J., Kim, J. & Yeom, H. W., Detecting photoelectrons from spontaneously formed excitons. *Nat. Phys.* **17**, 1024-1030 (2021). doi:10.1038/s41567-021-01289-x

36. Rajasekaran, S., Okamoto, J., Mathey, L., Fechner, M., Thampy, V., Gu, G. D. & Cavalleri, A., Probing optically silent superfluid stripes in cuprates. *Science* **359**, 575-579 (2018). doi:10.1126/science.aan3438

37. Matsunaga, R., Tsuji, N., Fujita, H., Sugioka, A., Makise, K., Uzawa, Y., Terai, H., Wang, Z., Aoki, H. & Shimano, R., Light-induced collective pseudospin precession resonating with Higgs mode in a superconductor. *Science* **345,** 1145-1149 (2013). doi:10.1126/science.1254697

38. Kabanov. V. V., Demser, J., Podonik, B. & Mihailovic, D., Quasiparticle relaxation dynamics in superconductors with different gap structures: theory and experiments on $YBa_2Cu_3O_{7-\delta}$. *Phys. Rev. B* **59**, 1497 (1999). doi:10.1103/PhysRevB.59.1497

39. Kabanov. V. V., Demser, J. & Mihailovic, D., Kinetics of a superconductor excited with a femtosecond optical pulse. *Phys. Rev. Lett.* **95**, 147002 (2005). doi:10.1103/PhysRevLett.95.147002

40. Michael, M. H., Först, M., Nicoletti, D., Haque, S. R. U., Zhang, Y., Cavalleri, A., Averitt, R. D., Podolsky, D. & Demler, E., Generalized Fresnel-Floquet equations for driven quantum materials. *Phys. Rev. B* **105**, 174301 (2022). doi:10.1103/PhysRevB.105.174301

**Data availability** The data presented in this manuscript are available from the corresponding author upon reasonable request.

**Acknowledgements** We thank D. Hsieh, P. Narang, M.K. Liu, and A. Kogar for fruitful discussions. S.R.U.H., M.H.M., J.Z., Y.Z., J.P.W., G.F.Z., J.Z., J.G.C., E.D. and R.D.A. acknowledge support from DARPA 'Driven Nonequilibrium Quantum Systems' (DRINQS) program under award no. D18AC00014. L.W., S.L. and A.R. acknowledge support from the European Research Council (ERC-2015-AdG694097), the Cluster of Excellence 'Advanced Imaging of Matter' (AIM), Grupos Consolidados (IT1249-19) and Deutsche Forschungsgemeinschaft (DFG) – SFB-925 – project 170620586. A.R. also thanks the Flatiron Institute, a division of the Simons Foundation.

**Author contributions** R.D.A. conceived the project together with S.R.U.H. J.Z., J.P.W. and J.G.C. performed the material growth and characterization. S.R.U.H. and J.Z. built the experimental setup. S.R.U.H., Y.Z. and G.F.Z. performed the optical pump – THz probe measurements. S.R.U.H. analyzed the data. M.H.M. and E.D. performed the first-principles calculation and numerical simulations. L.W., S.L. and A.R. performed the DFT analysis. All authors participated in the discussion and interpretations of the results. R.D.A. and E.D. supervised the project. S.R.U.H. and R.D.A wrote the manuscript with input from all authors.

**Competing interests** The authors declare that they have no competing interests.

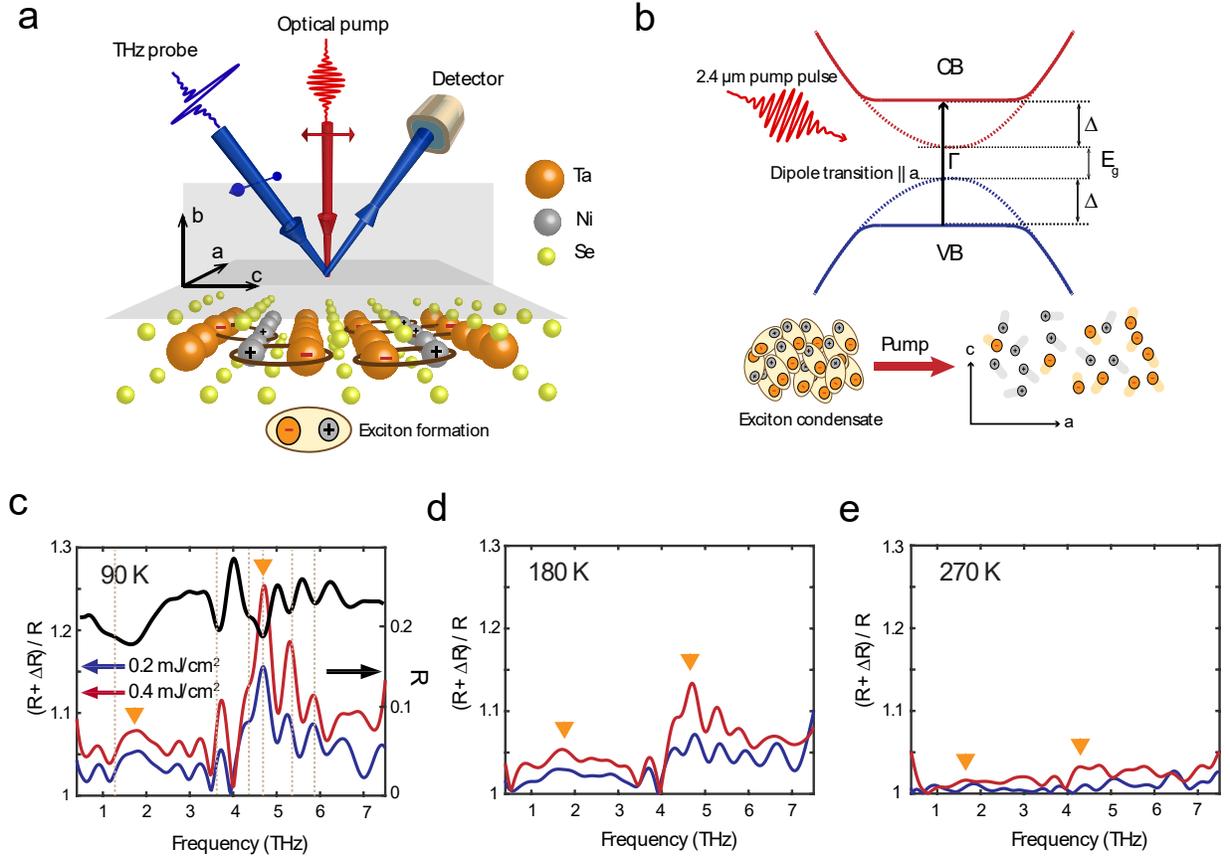

**Fig. 1| Experimental study of $Ta_2NiSe_5$. a**, Structure of $Ta_2NiSe_5$ (TNS) – alternating chains of Ta and Ni are aligned along the $a$-axis, and schematic of 2.4 μm pump (0.5 eV) THz probe experiment. Charge transfer between adjacent Ta and Ni chain forming excitons across the chain (represented with brown ellipse). **b**, Band structure of TNS showing the excitonic gap 2Δ, the dashed lines represent the band structure at $T > T_C$. The pump pulse is parallel to the $a$-axis. **c,** Photoinduced enhancement of reflectivity $(R + \Delta R)/R$ at 90 K as a function of fluence, 1 ps after photoexcitation (left panel). Solid blue curves are for a pump fluence of 0.2 mJ/cm² while red curves display 0.4 mJ/cm² results. Right panel shows equilibrium reflectivity at 90 K (black curve) with phonon locations in the spectrum denoted by dashed vertical gray lines. **d – e**, $(R + \Delta R)/R$ data for 180 K and 270 K, respectively. Orange inverted triangles serve as a guide to the eye to demonstrate the peaks in the reflectivity enhancement.

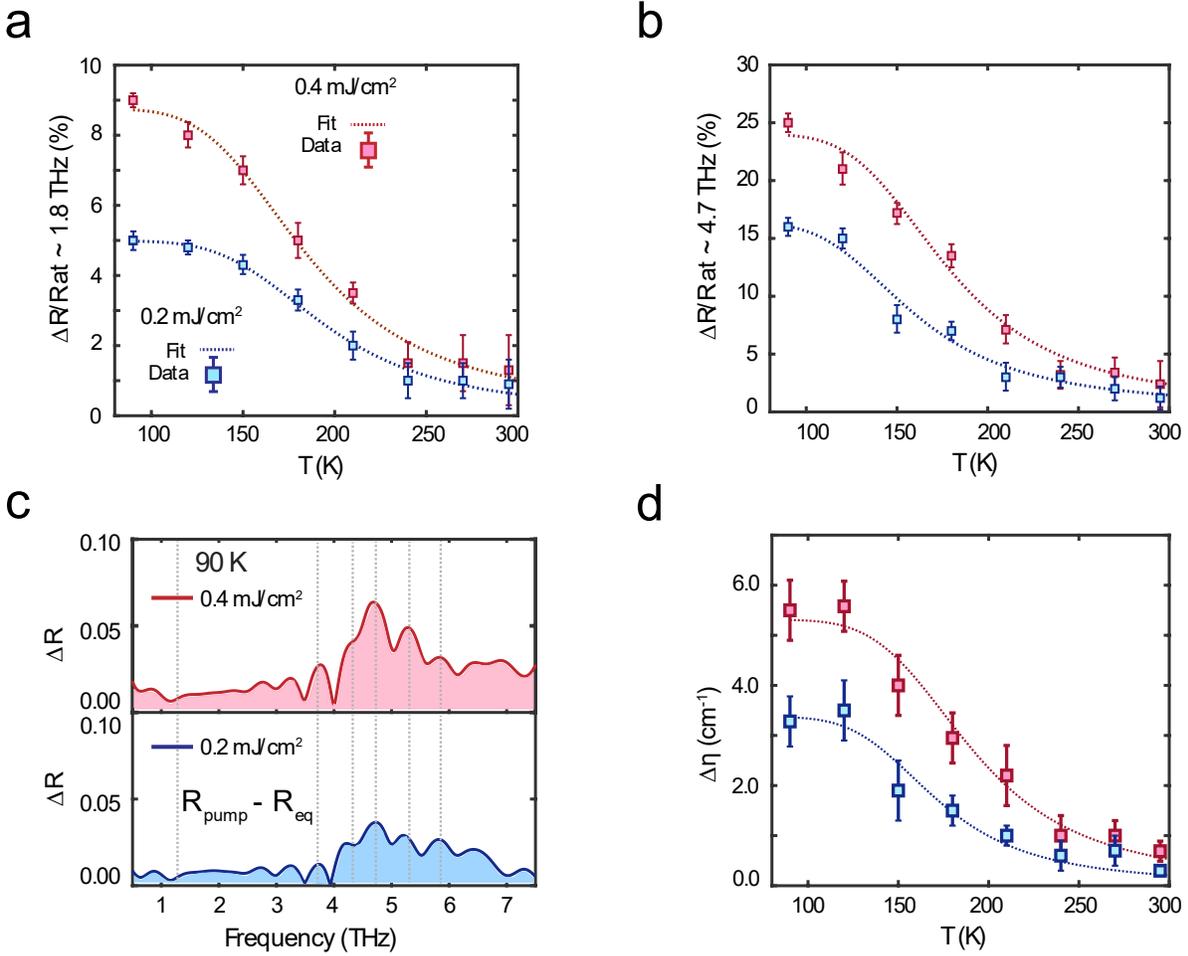

**Fig. 2| Temperature dependence of the reflectivity enhancement. a, b**, Temperature dependence of % change in reflectivity $\Delta R/R$ at ~1.8 THz and ~4.7 THz for both 0.2 mJ/cm$^2$ (solid blue squares) and 0.4 mJ/cm$^2$ (solid red squares) and the corresponding fits (dashed lines). Error bars represent standard deviations in fit values. **c**, Photoinduced change in reflectivity $\Delta R$ for excitation fluence of 0.4mJ/cm$^2$ (top panel) and 0.2mJ/cm$^2$ (bottom panel). The shaded regions represent the integrated change in the reflectivity. Phonon locations are indicated by vertical gray dashed lines as a guide to the eye. **d**, Temperature dependence of photoinduced change in integrated reflectivity $\Delta\eta(T)$ for each fluence. Dashed lines are fits as described in the text.

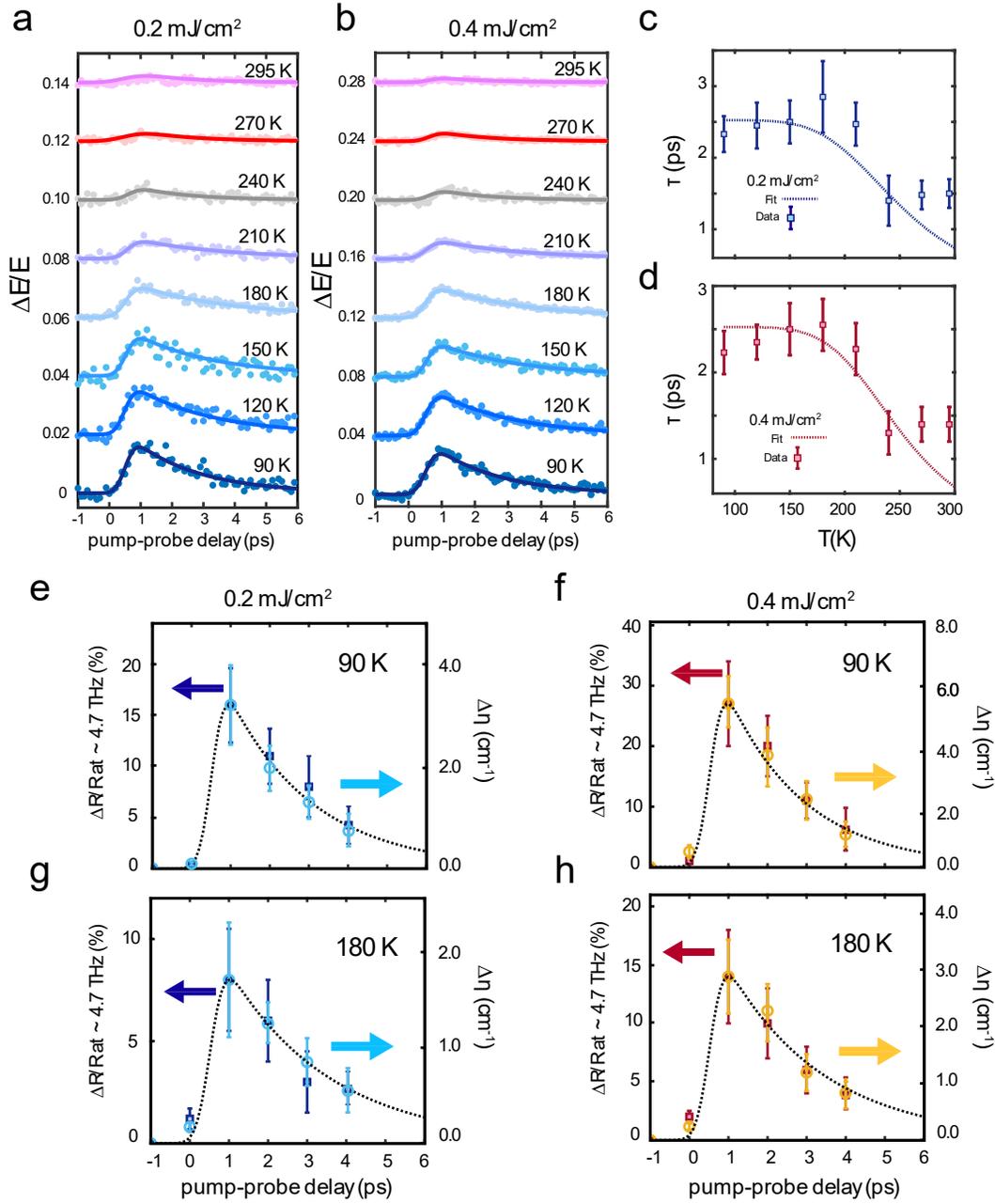

**Fig. 3| Reflectivity enhancement dynamics. a, b**, Photoinduced changes in peak THz probe electric field (solid circles) as a function of pump-probe delay time for $0.2\ \text{mJ/cm}^2$ (**a**) and $0.4\ \text{mJ/cm}^2$ fluence (**b**). Solid lines indicate a single-exponential decay fit. **c, d**, Temperature dependence of decay time as a function of fluence. Error bars represent the 95% confidence region from the single-exponential fits. **e – f**, Temporal evolution of $\Delta R/R(T)$ (left panel, solid squares) and $\Delta\eta(T)$ (right panel, open circles) at different pump-probe delays, as a function of fluence for 90 K and **g, h**, 180 K. Error bars represent maximum uncertainties from the fits.

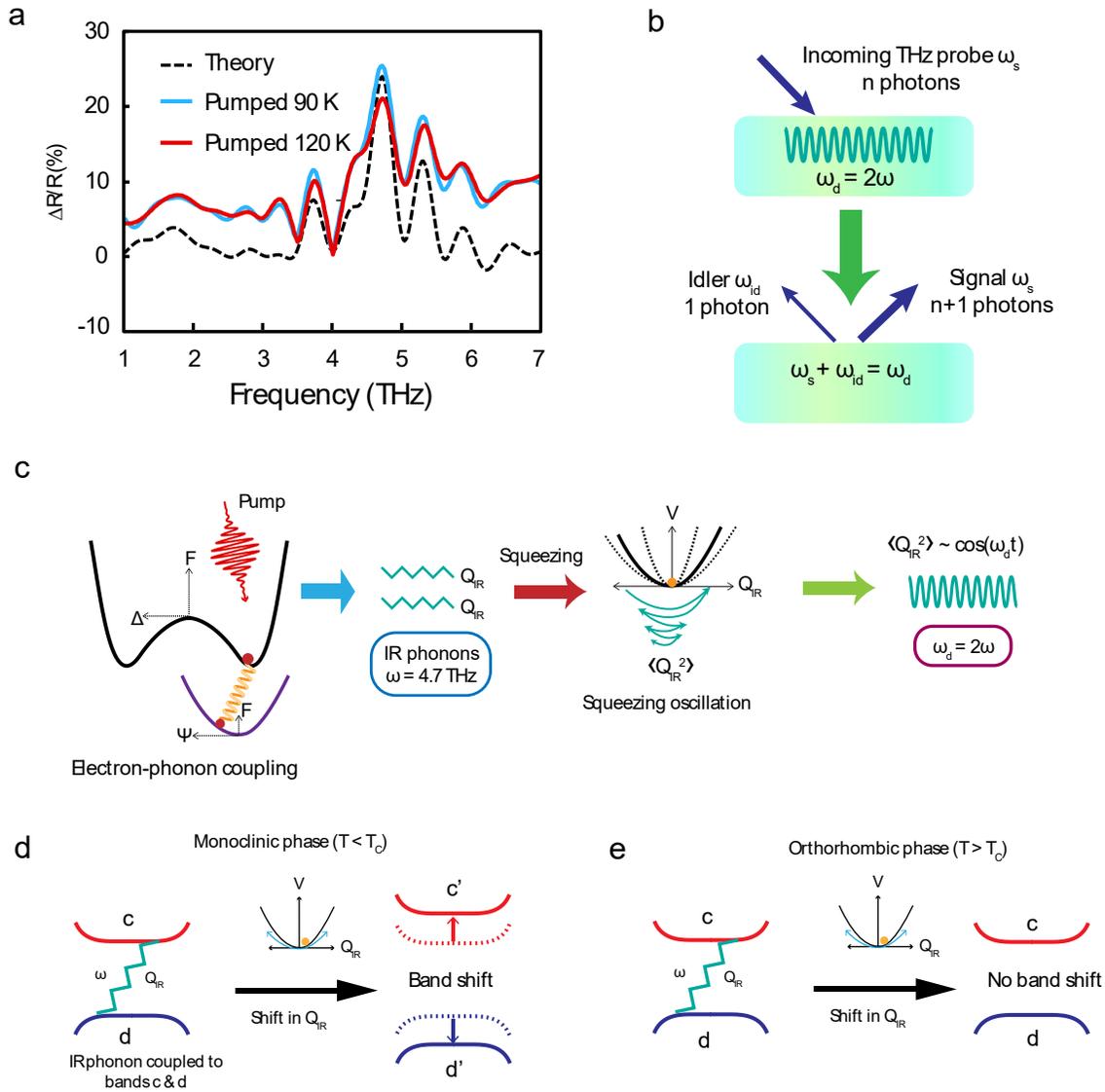

**Fig. 4| Description of parametric amplification at 4.7 THz. a**, Theoretical analysis (black dashed line) using Fresnel-Floquet analysis and experimental data (red and blue) showing good agreement. **b**, Incoming THz photon $\omega_s$ interacting with $2\omega$ drive and creating signal ($\omega_s$) and idler ($\omega_{id}$) photons, resulting in enhanced reflected signal at $\omega_s$. **c**, Photoinduced modulation of electron-phonon coupling for dominant 4.7 THz phonon. Squeezed phonons oscillating at frequency $2\omega$, act as the drive for parametric photon generation. **d**, Schematic summary of DFT calculations for 4.7 THz phonon revealing strong electron-phonon coupling resulting in a band shift in the monoclinic phase. **e**, In the high-temperature orthorhombic phase, no significant band shift arises from electron phonon coupling.

# Methods

**Sample preparation and characterization**

Single crystals of $Ta_2NiSe_5$ (TNS) were grown by a vapor transport method with iodine as the transport agent. Stoichiometric amounts of Ta, Ni, and Se powders were first pre-reacted at 900 °C before the vapor transport process, which was performed across a temperature gradient spanning 835 °C to 850 °C. The resultant crystals had lateral dimensions of 1 mm × 0.6 mm along the $a$ and $c$ axes, respectively, with a thickness of ~80 μm. The structure and single-phase purity of the resulting crystals was confirmed by powder X-ray diffraction. Extended Data Fig. 1 displays the temperature-dependent resistivity of TNS. A sudden drop in resistivity is observed at 326 K (indicated by the blue triangle), indicating a phase transition.

**Optical pump – broadband THz probe measurement**

Extended Data Fig. 2a shows a schematic of our experiment. For our near-infrared optical pump – broadband THz probe experiment, we used a regenerative amplified Ti:sapphire laser system with pulse energy of 3 mJ, pulse duration of 35 fs, 1 kHz repetition rate and 800 nm central wavelength as the main light source. The light source is split into three ways: one beam for the pump pulse and two beams for the generation and detection of the THz pulse. Horizontally polarized 2.4 μm pump pulses were generated after 1mJ 800 nm laser pulse was injected into a commercial TOPAS-C optical parametric amplifier (OPA). The beam diameter was 3 mm FWHM. The pulse width out of the OPA, 2 mirrors before the sample was measured to be 45 fs. 2.4 μm (0.5 eV) pump was chosen for 2 reasons: (1) this was the lowest pump energy in our pump-probe setup, and (2) it is the closest energy to the insulating gap of TNS measured by ARPES and FTIR measurements (0.16 — 0.18 eV)[4,8].

For the broadband THz generation, 800 nm pulses go through a type-I $\beta$-Barium Borate (BBO) crystal which generates vertically polarized second harmonic (400 nm) waves[41,42]. Then the fundamental and the second harmonic signals are passed through a 40 μm dual wave plate (DWP) which rotates the polarization of the 800 nm and 400 nm waves by 90° and 180° respectively. DWP behaves as a quarter wave plate (QWP) and half wave plate (HWP) with respect to 800 nm and 400 nm wavelengths, respectively. As such, both 800 nm and 400 nm waves become vertically polarized. Then both beams are focused into air to generate an air-plasma that radiates vertically polarized THz waves. The incident angle for the THz probe on the sample is 30° and the beam diameter is ∼1 mm. To detect the THz beam, we employed electro-optic sampling (EOS). For this purpose, we used a 300 μm-thick ⟨110⟩ GaP detection crystal that enabled spectral measurements from 0.5 – 7.5 THz. Extended Data Fig. 2b displays a detailed schematic of our broadband THz generation setup. We also show a THz time-domain signal from a gold mirror used as a reference and the corresponding Fourier transform amplitude (Extended Data Fig. 3a, 3b).

We used a 1 MHz bandwidth Newport 2307 balanced photodetector to detect the signal. For the data acquisition, a NI PXIe-5122 Data Acquisition (DAQ) system with 100 MHz maximum sampling rate and 100 MHz bandwidth was used. The DAQ was triggered by the output from an optical chopper used to modulate the pump pulse (For details, see Ref. [43]).

The TNS and gold reference samples were mounted inside a Janis ST-300 cryostat with a 36 mm-diameter, 600 μm-thick diamond window custom-made by Diamond Materials GmbH®. Integrating the diamond window to the cryostat gave us access to the broad THz spectrum at low temperatures owing to the broadband transmission properties of diamond. Since the pump and probe are horizontally and vertically polarized respectively, we oriented the sample in a way that the $c$-axis is along the horizontal direction, while the $a$-axis is oriented vertically. This setup enables us to photoexcite the sample along the $c$-axis and observe the dynamics along the $a$-axis (details shown in Fig. 1a).

**Equilibrium data**

To obtain equilibrium reflectivity of the sample, we measured $E_{gold}(t)$ and $E_{sample}(t)$. Here $E_{gold}(t)$ and $E_{sample}(t)$ are the reflection of the incident probe field from the gold reference and the TNS sample respectively, with $t$ being the electro-optic gate time. Fast-Fourier transforming (FFT) the signals yield the frequency-domain $E_{gold}(\omega)$ and $E_{sample}(\omega)$. The frequency-dependent reflectivity $R$ is given as: $R = |r|^2 = |E_{sample}(\omega) / E_{gold}(\omega)|^2$. Extended Data Fig. 4a, b display the temperature-dependent equilibrium $a$-axis reflectivity and optical conductivity of TNS, respectively. From the figures, it is clear that there are multiple IR-active phonons in TNS along the chain. The locations of the phonons are consistent with previous infrared measurements[8,28].

It was also observed that at lower temperatures, the conductivity displays sharper peaks, and with increasing temperature, the sample exhibits larger optical conductivity values. As we approach the room temperature, the conductivity starts developing flat features, similar to that in ref. [8], in harmony with the equilibrium reflectivity data.

**Pump-probe data**

We measured the pump-induced transient change $\Delta E(t, t_{pp})$ in the reflected electric field of the probe pulse as a function of pump-probe delay time $t_{pp}$ over the THz temporal window $t$. The signal $\Delta E(t, t_{pp})$ can be expressed as, $\Delta E(t, t_{pp}) = E_{sample,pumped}(t, t_{pp}) - E_{sample,unpumped}(t, t_{pp})$. From this, the total complex reflection coefficient, $r(\omega, t_{pp}) =$ $r_0(\omega)[1 + \frac{\Delta E(\omega, t_{pp})}{E_{sample,unpumped}(\omega, t_{pp})}]$ is determined[44], where $r_0(\omega)$ is the equilibrium reflection coefficient previously obtained from equilibrium measurement. To measure the transient THz signal without any artifacts, the standard Kindt and Schmuttenmaer scanning scheme was used[44,45].

To determine the complex optical parameters of the photoexcited region, we need to also consider the penetration depth mismatch of the 2.4 µm pump (~250 nm) and the THz probe (50 – 300 µm). Integrating the photoexcited change in reflectivity $\Delta R(\omega)$ over the total spectral range in our experiment (0.5 – 7.5 THz), the total pump-induced reflectivity spectral weight change $\Delta \eta$ was further extracted. Using this formalism, we also determine the reflectivity enhancement $(R + \Delta R)/R$ as well as the pump-induced percent change in reflectivity $\Delta R/R$.

Extended Data Fig. 5 displays the temperature and fluence-dependent dynamics of the reflectivity enhancement $\Delta R/R$ (left panel, closed squares) and photoinduced change in spectral weight $\Delta \eta$ (right panel, open circles) at different $t_{pp}$. All the data were plotted on the same scale in order to facilitate a detailed comparison of the dynamics. The black dashed lines denote single exponential fits using the function, $f(t_{pp}) = a \left( \text{erf}\left(b(t_{pp} + c)\right) + 1 \right) e^{-\frac{t_{pp}+c}{\tau}}$. Here, $a, b$ and $c$ are fitting parameters while $\tau$ denotes the relaxation time. As evident from the figures, $\Delta R/R$ and $\Delta \eta$ exhibit similar dynamics which verifies our observation that the reflectivity enhancement is attributed to an electronic origin arising from the photoexcitation in TNS. It was also noted that the dynamics of both $\Delta R/R$ and $\Delta \eta$ closely match the dynamics shown in Fig. 3a, 3b.

**Specific heat analysis**

We estimate the lattice temperature after photoexcitation. For a given pump fluence, we estimate the absorbed energy over the pumped volume and calculate the effective temperature after electron-phonon thermalization using the relation, $Q_{pump} = \int_{T_i}^{T_f} N\, C(T) dT$, where $Q_{pump}$ is the total absorbed energy, $N$ is the number of moles in the excited volume, $C(T)$ is the temperature-dependent lattice specific heat of TNS, and $T_i$ and $T_f$ are the initial and final temperature after electron phonon thermalization, respectively. $Q_{pump}$ is derived from the relation, $Q_{pump} = F A (1 - R)$. Here $F$ is the pump fluence, $A$ is the area of the pump beam on the sample with a diameter ~3 mm (pump beam FWHM diameter), and $R$ is the reflectivity at 2.4 μm (0.5 eV) which is ~0.45[30]. $N$ can be written as $N = V_{excited}/V_{molar}$, where photoexcited volume $V_{excited}$ is estimated as a cylindrical disk with a diameter of ~3 mm and height equal to the penetration depth $d$, and molar volume $V_{molar}$ was calculated as 106 cm$^3$/mol. We used the temperature dependent specific heat $C(T)$ data from Ref. [6] and fitted it using Debye model (Extended Data Fig. 6a).

The final temperature was calculated numerically and plotted as a function of fluence. Extended Data Fig. 6b displays temperature raise $\Delta T$ as a function of fluence for different initial temperatures $T_i$. We note that, upon photoexcitation the temperature change is very small; even for the highest fluence in our experiment (0.4 mJ/cm$^2$), the temperature rise is only ~5 K for an initial temperature of 90 K and ~3 K for an initial temperature of 300 K. This eliminates the possibility of any significant thermal heating effect and further confirms that our experimental findings are largely due to electronic (excitonic) correlations. Since the insulator-semiconductor phase transition temperature for TNS is 326 K, the specific heat analysis also shows that the sample remains in excitonic insulator phase (below $T_c$) after photoexcitation.

**Parametric reflectivity amplification**

In this section we provide additional details of the phenomenological Fresnel-Floquet analysis. To compute the material reflectivity at the probe pulse frequency, a better understanding of the character of light propagation inside the material is needed. It requires solving Maxwell equations with the frequency-dependent refractive $\tilde{n}(\omega)$ to find the wavevector of light $k$ inside the medium. This satisfies the equation,

$$\left(\frac{\tilde{n}(\omega)^2 \omega^2}{c^2} - k^2\right) E(\omega) = 0. \tag{1}$$

Here $E(\omega)$ is the electric field of the probe. Boundary conditions at the interface of the material for electric and magnetic fields lead to the usual Fresnel equations which relates $\tilde{n}(\omega)$ to the complex reflection coefficient $r(\omega)$. For normal incidence, it has the simple form,

$$\tilde{n}(\omega) = \frac{1 - r(\omega)}{1 + r(\omega)}. \tag{2}$$

A similar approach can be used to in the presence of an oscillating field at frequency $\omega_d$. We model the effects of the oscillating field on the photons phenomenologically through an oscillating permittivity:

$$\delta\varepsilon(t) = 2A_d \cos(\omega_d t). \tag{3}$$

This gives rise to a parametric drive with frequency $\omega_d$ and a subsequent mixing of signal ($\omega_s$) and idler ($\omega_{id}$) frequency components, satisfying the relation $\omega_s + \omega_{id} = \omega_d$ (For more details, see Ref. [46]). Hence, light eigenmodes inside the material should be understood as the Floquet states that take the form,

$$E(t) = e^{ikx}\left(E_s e^{-i\omega_s t} + E_{id} e^{i\omega_{id} t} + c.c.\right), \tag{4}$$

where the wavevector $k$ was found from the solving the Maxwell equations. Using (1), (3) and (4) the new equations in the presence of the parametric drive term are,

$$\left(\frac{n(\omega_s)^2\omega_s^2}{c^2} - k^2\right)E_s(\omega) + A_d E_{id} = 0, \tag{5a}$$

$$\left(\frac{n(\omega_{id})^2\omega_{id}^2}{c^2} - k^2\right)E_{id}(\omega) + A_d E_s = 0. \tag{5b}$$

Because of the coupling between signal and idler components, every frequency will have two transmission channels associated with different wavevectors, and two eigenmodes inside the material for an incoming field $E_0$:

$$E_j = t_j E_0 e^{ik_j x}\left(e^{-i\omega_s t} + \alpha_j e^{i\omega_{id} t}\right), \tag{6}$$

where $j = 1, 2$, $\alpha_j$ is the relative amplitude given by the eigenvector with $k_j$, while $t_j$ is the transmission coefficient of the j-th channel. The reflectivity is computed by solving the Fresnel-Floquet equations which involve matching electric and magnetic fields parallel to the sample surface for both signal and idler frequencies. Inside the sample the total electric field becomes,

$$E_{sample} = \sum_{j=1}^{2} t_j E_0 e^{ik_j y}\left(e^{-i\omega_s t} + \alpha_j e^{i\omega_{id} t}\right), \tag{7}$$

and the electric field in the air becomes,

$$E_{air} = E_0\left(e^{\frac{i\omega_s y}{c} - i\omega_s t} + r_s e^{-\frac{i\omega_s y}{c} - i\omega_s t} + r_{id} e^{\frac{i\omega_{id} y}{c} + i\omega_{id} t}\right). \tag{8}$$

Here $r_s$ and $r_{id}$ denote the complex reflection coefficient at signal and idler frequencies, respectively. Using Maxwell's equations, the magnetic field and matching boundary condition are determined at $y = 0$ (sample-air interface). This yields the equations for the driven system:

$$1 + r_s = t_1 + t_2, \tag{9a}$$

$$1 - r_s = \frac{k_1 t_1}{\omega_s} + \frac{k_2 t_2}{\omega_s}, \tag{9b}$$

$$r_{id} = t_1 \alpha_1 + t_2 \alpha_2, \tag{9c}$$

$$r_{id} = \frac{k_1 t_1 \alpha_1}{\omega_{id}} + \frac{k_2 t_2 \alpha_2}{\omega_{id}}. \tag{9d}$$

From these equations, it is possible to determine the complex reflection coefficients at signal and idler frequencies. To fit the data, the complex refractive index is determined from the equilibrium reflectivity. A 9.4 THz oscillation in the permittivity is used as the primary drive providing a time-dependent transient change in the optical properties of the sample upon photoexcitation. Combining the drive with the equilibrium properties of the sample yields the photoexcited optical parameters along with the photoexcited reflectivity. The fit of $\Delta R/R$ is shown in Fig. 4a which shows a very good agreement between experiment and theory. However, that the experimental data has a larger value than the theoretical fit.

**Theory of parametrically amplified reflectivity in TNS**

Optical pumping at 0.5 eV excites electron-hole pairs and is off-resonant with respect to the phonons. From symmetry arguments, the photo-excited electron occupation couples to IR active modes through a phonon squeezing Hamiltonian:

$$H = \sum_k g_k \, Q_{IR}^2 \, c_k^\dagger c_k. \tag{10}$$

In (10), $g_k$ is the electron-phonon coupling strength and $Q_{IR}$ is the IR phonon coordinate. It can be shown how such a term can naturally arise through a dipole-dipole interaction between IR active phonon dipole moments linearly coupled to electron dipole transitions. Squeezing of IR active phonons by photo-excited electrons through this type of interaction has been explored in previous studies in the context of photo-induced superconductivity[47]. We use ab-initio DFT calculations outlined below to calculate $g_k$ and find a very strongly coupled IR phonon at 4.7 THz.

**Phonon squeezing**

In this section a detailed account of the IR-phonon squeezing process and the coherent oscillation at twice the phonon frequency is presented. Phonon squeezing provides a route to generate phonons at frequency $\omega$ with higher energy (i.e., $> \omega$) pump photons. In this case, the expectation value of the phonon coordinate is zero (i.e., $\langle Q_{IR} \rangle = 0$). However, phonon fluctuations can be excited through a Raman process, $\langle Q_{IR}^2(t) \rangle = \langle Q_{IR}^2 \rangle_0 + A\cos(2\omega t)$. To clarify this, the effective Hamiltonian for a phonon system can be written as,

$$H_{ph} = \frac{\Pi^2}{2M} + \frac{M(\omega_0^2 + f(t))Q_{IR}^2}{2} + ZEQ_{IR}, \quad (11)$$

where $\Pi$ is the momentum of the phonon coordinate $Q_{IR}$, $\omega_0$ is the bare phonon frequency (in this case, it is 4.7 THz), $M$ is the phonon mass, $Z$ is the IR activity of the phonon, $E$ is the electric field, and the effective parametric drive $f(t)$ is given by the photoexcited electron population coupled to the phonon,

$$f(t) = \sum_k \frac{g_k}{M} n_k(t). \quad (12)$$

Here, $g_k$ is the electron-phonon coupling (as above) and $n_k(t)$ is the photoexcited electron occupation. The electron occupation dynamics was modeled as a fast photoexcitation process that thermalizes back to equilibrium on a characteristic timescale and the parametric drive $f(t)$ is written as $f(t) = f_0 \theta(t) e^{-\gamma t}$. If the characteristic thermalization time is of the order of ~2 ps, then $f(t)$ can be taken as a theta function. In contrast, for timescales much smaller than 100 fs, the period of phonon oscillations, $f(t)$ is a delta function. In both cases, fast electron dynamics provide an impulsive Raman drive with a broad frequency spectrum which linearly excites phonon fluctuations $\langle Q_{IR}^2(t) \rangle$. The equations of motions of the phonon field fluctuations are:

$$\partial_t \langle Q_{IR}^2 \rangle = \frac{\langle Q_{IR}\Pi + \Pi Q_{IR} \rangle}{M}, \tag{13a}$$

$$\partial_t \langle Q_{IR}\Pi + \Pi Q_{IR} \rangle = -2M(\omega_0^2 + f(t))\langle Q_{IR}^2 \rangle + 2\frac{\langle \Pi^2 \rangle}{M} - 2Z\langle E Q_{IR} \rangle, \tag{13b}$$

$$\partial_t \langle \Pi^2 \rangle = -M(\omega_0^2 + f(t))\langle Q_{IR}\Pi + \Pi Q_{IR} \rangle - Z\langle E\Pi + \Pi E \rangle, \tag{13c}$$

At $k = 0$, we substitute $E = \frac{ZQ_{IR}}{\varepsilon \varepsilon_0}$, and rewrite the equations (13b) and (13c) as,

$$\partial_t \langle Q_{IR}\Pi + \Pi Q_{IR} \rangle = -2M(\omega^2 + f(t))\langle Q_{IR}^2 \rangle + 2\frac{\langle \Pi^2 \rangle}{M}, \tag{13d}$$

$$\partial_t \langle \Pi^2 \rangle = -M(\omega^2 + f(t))\langle Q_{IR}\Pi + \Pi Q_{IR} \rangle, \tag{13e}$$

where $\omega^2 = \omega_0^2 + \frac{Z^2}{\varepsilon \varepsilon_0 M}$ is the renormalized phonon frequency at $k = 0$. Expanding the phonon fluctuations gives $\langle Q_{IR}^2 \rangle = \langle Q_{IR}^2 \rangle_0 + \langle Q_{IR}^2 \rangle_1$. Here $\langle Q_{IR}^2 \rangle_0$ is the thermal fluctuation and $\langle Q_{IR}^2 \rangle_1 \propto f(t)$. To linear order in $f(t)$, the equations of motion are $\frac{\langle \Pi^2 \rangle_1}{M} = M\omega^2 \langle Q_{IR}^2 \rangle_1 + Mf(t)\langle Q_{IR}^2 \rangle_0 + \mathcal{O}(f^2)$. Combining equations (13a), (13d) and (13e) yields:

$$(\partial_t^2 + 4\omega^2)\langle Q_{IR}^2 \rangle_1 = -4f(t)\langle Q_{IR}^2 \rangle_0. \tag{14}$$

Equation (14) provides insight about the IR phonon driving and the relevant phonon fluctuation frequency. It shows that parametric resonance of the phonon coordinate at twice the phonon frequency $2\omega$ appears as a regular resonance for the phonon fluctuations $\langle Q_{IR}^2 \rangle$[48]. It is also significant that the resonance of the phonon fluctuations acts as a filter for the broadband electron dynamics leading to coherent oscillations. These oscillations are responsible for parametrically amplifying the TNS reflectivity at select IR phonon frequencies. It can also be noticed that the left side of (14) resembles the simple harmonic oscillation equation of phonon fluctuation $\langle Q_{IR}^2 \rangle_1$ where the frequency of the oscillation ($2\omega$) is twice the phonon frequency $\omega$. This mathematical formulation demonstrates the source of the $2\omega$ driving frequency in the parametric amplification process and further justifies the excellent agreement between theory and experiment as depicted in Fig. 4a.

To show that this phenomenon is related to squeezing from a quantum mechanical perspective, we now expand $\langle Q_{IR}^2 \rangle$ in terms of creation and annihilation operators. Using the definition $Q_{IR} = \frac{a+a^\dagger}{\sqrt{2M\omega}}$, we find the expression,

$$\langle Q_{IR}^2 \rangle = \frac{\langle a^\dagger(t)a^\dagger(t) \rangle + \langle a(t)a(t) \rangle + \langle a^\dagger(t)a(t) \rangle + \langle a(t)a^\dagger(t) \rangle}{2M\omega}. \qquad (15)$$

Here, expectation values of $a(t)a^\dagger(t)$ and $a^\dagger(t)a(t)$ do not oscillate rapidly while the anomalous pairs $a^\dagger(t)a^\dagger(t)$ and $a(t)a(t)$ oscillate at twice the phonon frequency $2\omega$. As a result, a state with phonon fluctuation that oscillate at twice the phonon frequency implies the existence of a condensate of phonon pairs $\langle a^\dagger(t)a^\dagger(t) \rangle \neq 0$. This creation of anomalous phonon pairs is associated to the squeezing process.

Thus, our temperature dependent reflectivity enhancement data, together with this mathematical analysis, and phenomenological modeling establishes the phonon squeezing as the dominant effect, suggestive of a parametric process.

**Possibility of thermal phonon shift as an alternative interpretation**

There are two experimental signatures that favor a parametric amplification scenario: (1) as a function of frequency, the photoinduced change in the reflectivity is positive owing to energy extracted from the oscillating drive inside the material into the reflected beam[40,49], and (2) it is centered around 4.7 THz in a symmetric way. This indicates that parametric amplification arising from Raman-type modulation at 9.4 THz is the most natural mechanism that fits the data. To further verify, we considered other plausible scenarios that might support the experimental observation. As a result, we took into account the possibility of thermal phonon shifting as a competing mechanism.

Extended Data Fig. 7a shows equilibrium reflectivity at 90 K (black curve) as well as the reflectivity red shifted by 0.1 THz (red dashed curve), assuming that thermal effects red shifts the reflectivity. If thermal phonon shift was responsible for the reflectivity enhancement, the difference between equilibrium reflectivity and thermally red shifted reflectivity should ideally match $\Delta R$, the pump-induced change in reflectivity. As seen in Extended Data Fig. 7b, there is an unambiguous difference in trend between the photoinduced reflectivity change (red curve) and the reflectivity change due to red shift of phonons (orange curve). It was noted that, thermal shift of phonons results in both negative and positive changes in reflectivity. Whereas according to Fresnel-Floquet parametric amplification theory, an overall positive change in reflectivity is expected.

In addition, blue shift of phonons was also considered (Extended Data Fig. 7c), and as depicted in Extended Data Fig. 7d, negative and positive changes in reflectivity were observed similar to the red shifted case. It is evident that there is no similar pattern between photoinduced reflectivity enhancement (red curve) and reflectivity change due to blue shifted phonons (purple curve). This is contradictory to our experimental observation of a universally positive value $\Delta R$ as a function of frequency. However, this phenomenon is well captured by the parametric amplification formalism which proves to be the most natural explanation for the data.

**Possibility of THz emissions**

Possibilities of THz emission from the condensate were considered as another alternate scenario. To confirm whether there is THz emission, the THz probe was blocked to see if the detector picked up any THz signal emitted from the sample due to photoexcitation. However, no THz emission was observed. TNS is a centrosymmetric crystal with inversion symmetry, resulting in the exciton condensate having inversion symmetry as well. As a result, radiation is not expected by optical rectification method, as seen in ZnTe or $LiNbO_3$, without a current or magnetic field bias.

Extended Data Fig. 8 pictorially elucidates the excitation process. When pumped with an intense IR pulse, electron-phonon coupling triggers the squeezing of the IR-active phonons and results in the overtone (squeezing oscillation) at twice the phonon frequency. This overtone cannot emit radiation owing to the fact that it does not have any dipole (Extended Data Fig. 8a).

Rather we propose the stimulated THz emission scenario (Extended Data Fig. 8b), where the probe THz is amplified by the oscillating material parameters (phonon squeezing effect in our case) upon pumping. Although the $2\omega$ squeezing overtone has no dipole and hence, cannot emit THz, it can emit stimulated radiation. This can be explained by parametric amplification. First, the optical pump initiates the material dynamics at frequency $2\omega$ ($= \omega_d$) oscillation which is twice the IR phonon frequency $\omega$. Then the probe THz pulse reaches the sample. Since it is a broadband pulse, it consists of every frequency component from 0 to 7.5 THz (including $\omega$ that coincides with the IR phonon frequency). When the signal pulse frequency $\omega_s$ pulse interacts with the interface that is oscillating at $\omega_d$ frequency and satisfies the parametric resonance condition, it emits a pair of photons at $\omega_s$ and $\omega_{id} = \omega_d - \omega_s$, respectively. As we can see, from this stimulated emission process we get an extra photon at $\omega_s$ frequency, which results in the amplification of the signal picked up by the detector.

**DFT calculations**

The details of the DFT calculations are as follows: The phonon spectrum at the Γ point was computed using the Density Functional Perturbation Theory routines of the VASP code[50-53] employing the vdW-opt88 functional[54,55] on a $48 \times 4 \times 6$ mesh. The Born effective charges were calculated on a similar k-mesh while the IR-activity was derived using phonopy[56] and the phonon spectroscopy code by described in Ref. [57]. The electronic band structure plots have been obtained using a $24 \times 4 \times 6$ k-mesh and a 320 eV cutoff using the standard PBE functional.

For the monoclinic geometry there are 21 symmetry allowed IR-active modes, 11 $B_u$ and 10 $A_u$ modes. Among these, only four modes are in the frequency range near 4.5 THz, listed in Table 1. Upon displacing the phonon coordinates $Q_{IR}$ in the positive or negative direction about the equilibrium positions, we recalculated the electronic band structure. From the calculation, it is evident that only mode 26 (frequency ~4.7 THz) shows a strong modulation of the conduction bands upon a small displacement of the phonon coordinate, suggesting a strong coupling of this mode to the electronic degrees of freedom (Extended Data Fig. 9). The other phonons do not display significant coupling to the band structure. As a result, no shift in the band structure was observed for the phonon displacements along their eigenmodes. Since we observe the reflectivity enhancement at ~4.7 THz where there is a $B_{3u}$ phonon, the DFT calculation supports our experimental data as well as the theoretical formalism of parametric amplification. Extended Data

Fig. 10 displays the evolution of the electron-phonon coupling for the 4.7 THz mode for monoclinic (Extended Data Fig. 10a) and orthorhombic (Extended Data Fig. 10b) phases. It is shown from the *ab initio* calculation that the electron-phonon coupling for the 4.7 THz phonon is strong in the monoclinic phase. This results in a large band renormalization. On the contrary, the band shift almost disappears in the orthorhombic phase. This underscores the relation of the IR-active mode with the order parameter of TNS which is also evident in our experimental observations.

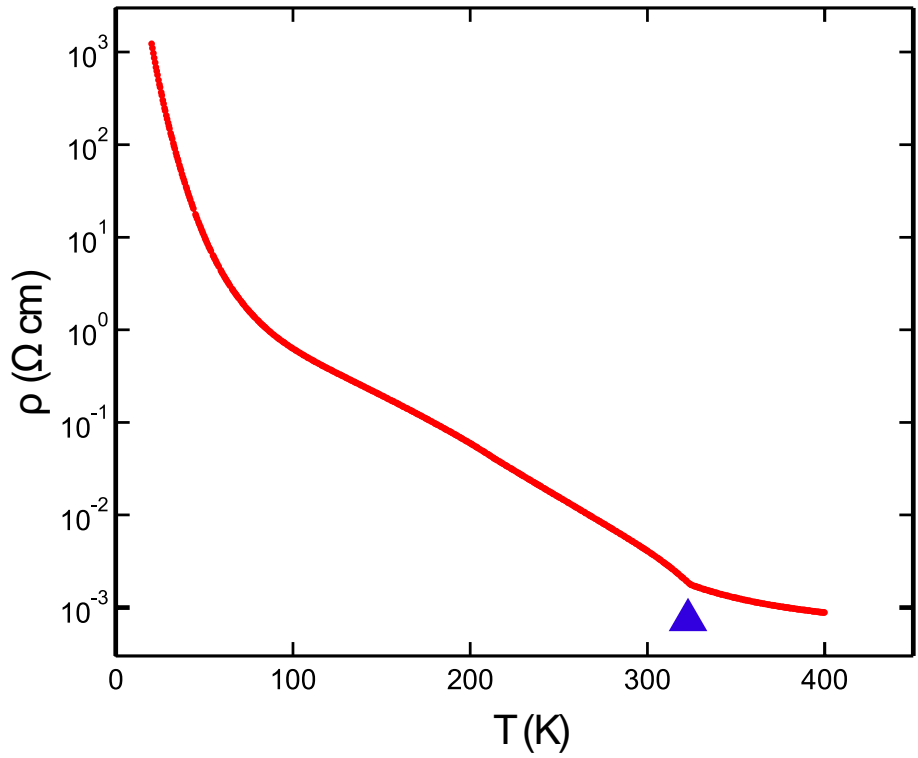

**Extended Data Fig. 1| Transport properties of $Ta_2NiSe_5$.** Resistivity of $Ta_2NiSe_5$ single crystal plotted against the temperature. The blue triangle emphasizes an anomaly at $T = 326$ K, corresponding to the temperature of the phase transition.

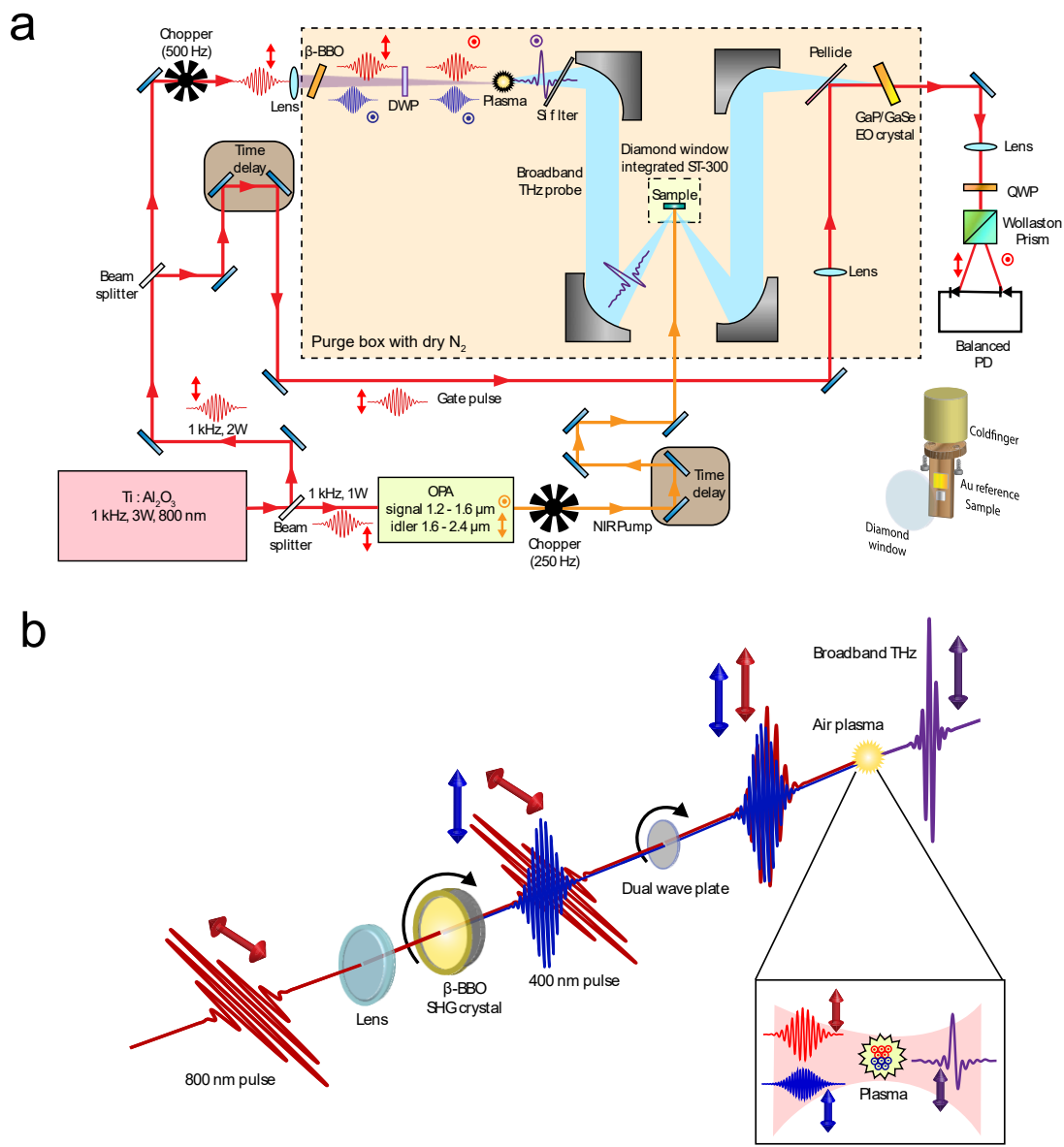

**Extended Data Fig. 2| Experimental setup and THz generation. a**, Experimental setup for near-infrared pump – broadband THz probe spectroscopy. DWP: dual wave plate, OPA: optical parametric amplifier, BBO: $\beta$-Barium Borate, QWP: quarter wave plate, PD: photodiode. 2.4 μm pump beams are generated as idler beams from the OPA while broadband THz probe pulses are generated from a two-color laser-induced air plasma. GaP crystal is utilized as the EO sampling crystal. **b**, Horizontally polarized 800 nm pulses generate vertically polarized 400 nm pulses after passing through the BBO crystal. The DWP rotates the 800 nm polarization from horizontal to vertical. Both beams are focused into the air and create an air plasma which radiates broadband THz waves. Arrows indicate the polarization directions. Inset shows a detailed schematic of the THz generation process from the air plasma.

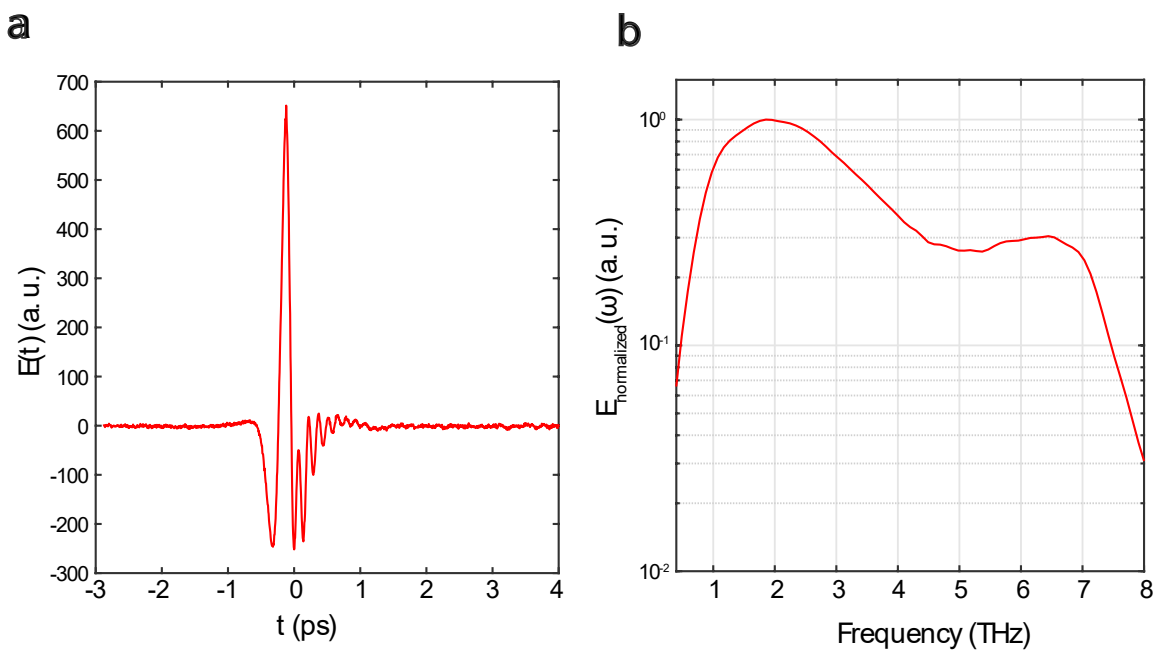

**Extended Data Fig. 3| Time-domain THz signal and corresponding spectrum. a**, Reflected time-domain signal (TDS) from a gold mirror (reference), a 300 μm-thick ⟨110⟩ GaP was used to detect the signal. **b**, The corresponding normalized spectrum showing a broadband spectral regime of 0.5 – 7.5 THz.

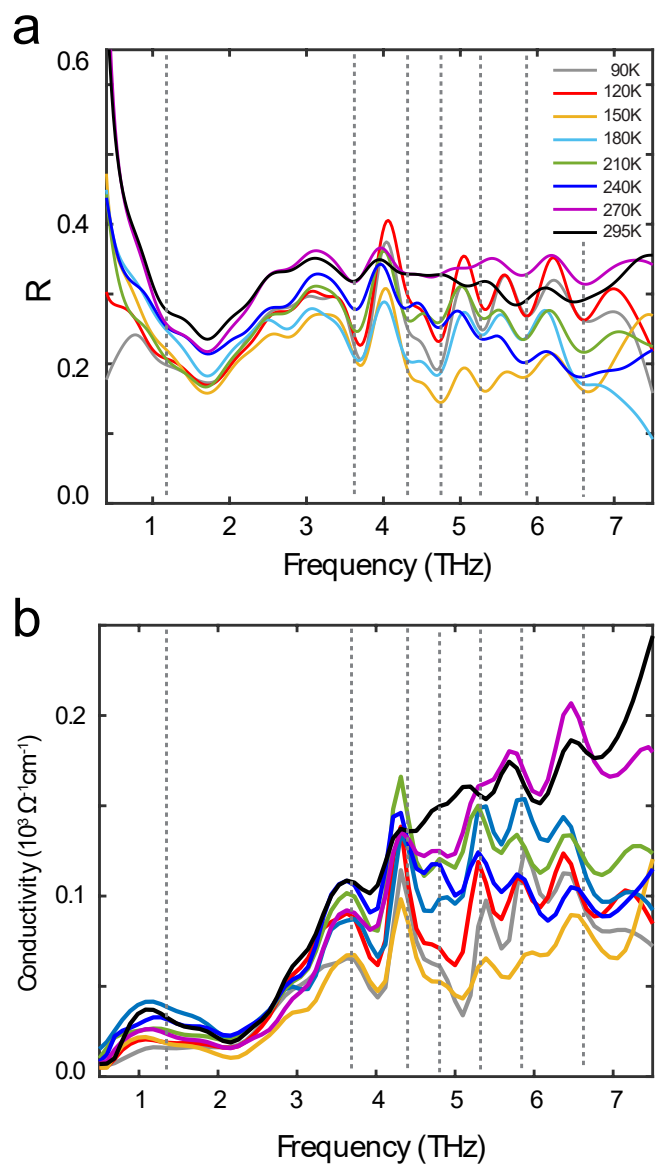

**Extended Data Fig. 4| Equilibrium reflectivity and optical conductivity. a**, Equilibrium reflectivity and **b**, optical conductivity along the *a*-axis as a function of temperature. Phonon locations are denoted by gray dashed lines.

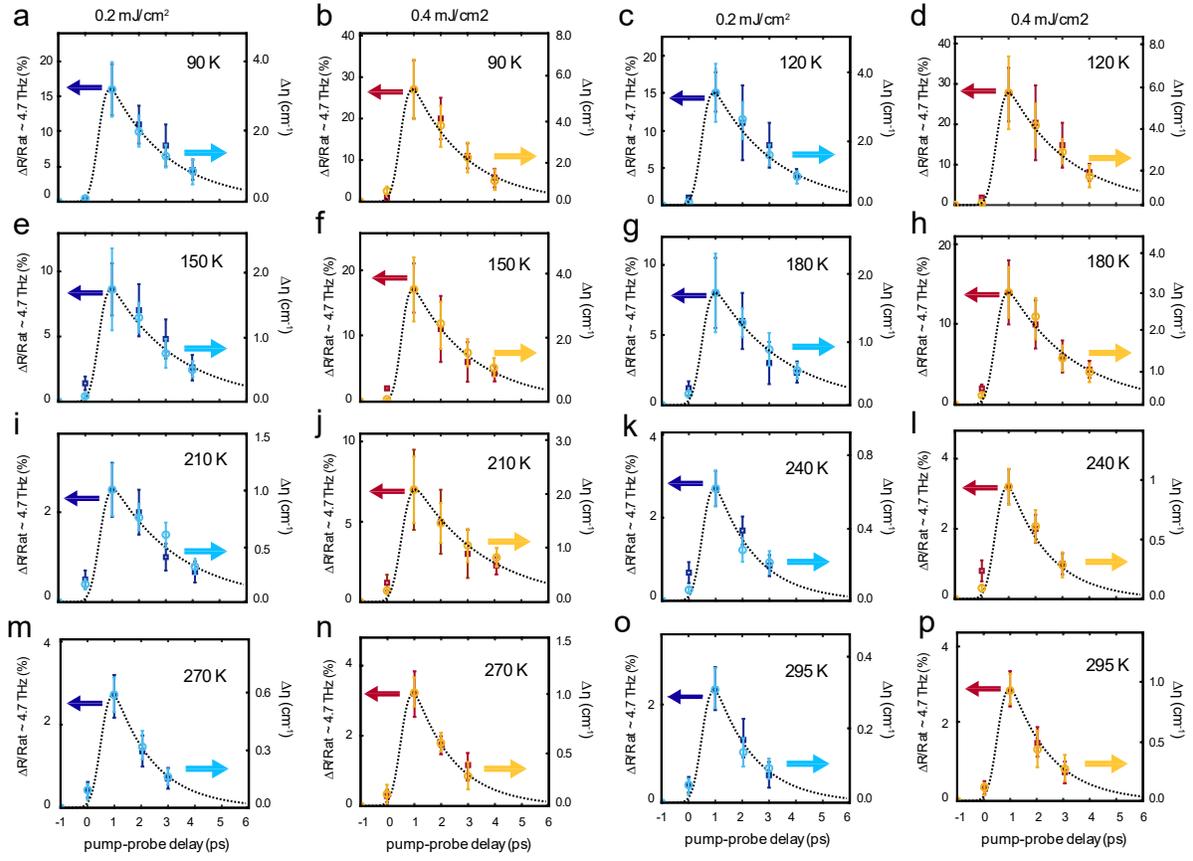

**Extended Data Fig. 5| Dynamics of the reflectivity enhancement.** Dynamics of reflectivity enhancement $\Delta R/R$ at 4.7 THz (left panel, closed squares) and integrated pump-induced change in reflectivity spectral weight $\Delta\eta$ (right panel, open circles) as a function of temperature and fluence. Both sets of data were plotted on the same scale for comparison. The dotted lines represent single-exponential decay function utilized to determine the relaxation time $\tau$. Error bars represent maximum uncertainties determined from the fits.

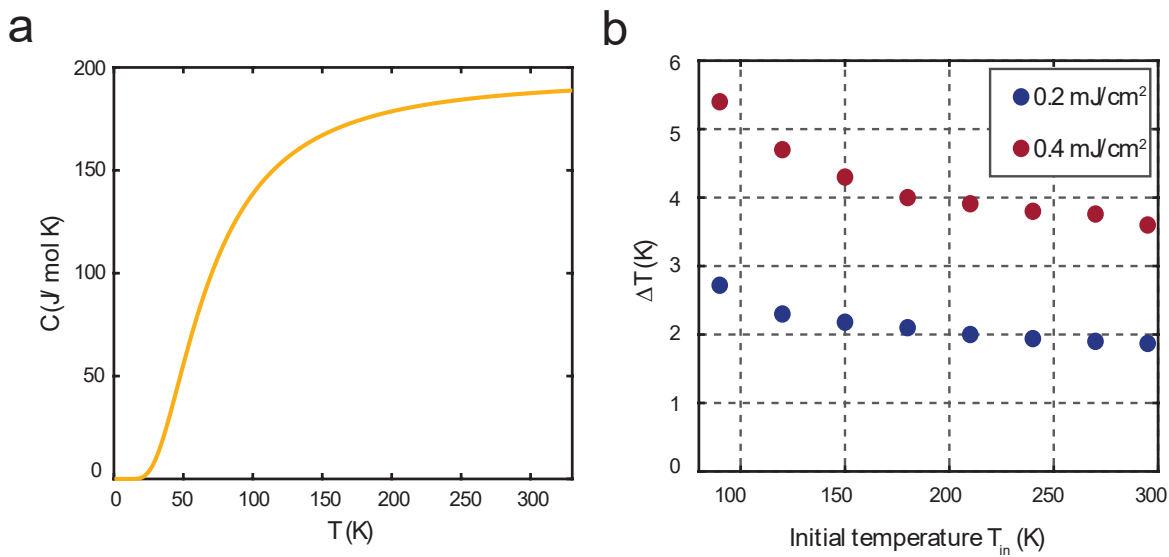

**Extended Data Fig. 6| Specific heat analysis. a**, Specific heat of TNS obtained from Ref. [6], then modeled with a Debye fitting. **b**, Fluence-dependent temperature increase $\Delta T$ for different initial temperatures $T_i$. It is observed from the plot that the photoinduced temperature rise is minimal, ruling out a thermal origin of the signal.

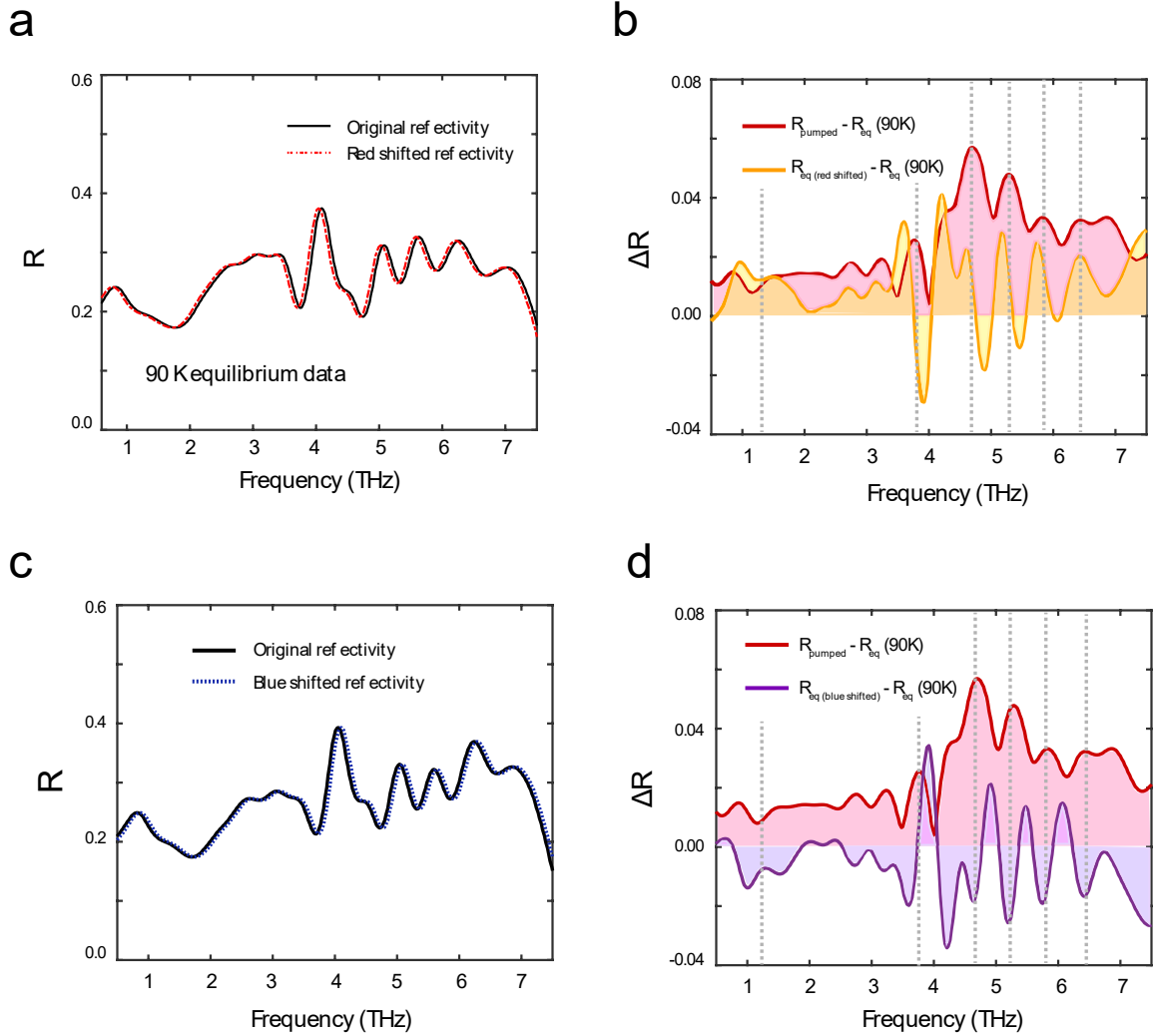

**Extended Data Fig. 7| Possibility of thermal phonon shift. a**, Equilibrium reflectivity at 90 K (black line) and reflectivity red shifted by 0.1 THz (red dashed line). **b**, Photoinduced reflectivity change $R_{pumped} - R_{eq}$ (red curve) along with reflectivity change due to red shift of phonons (orange curve) showing very different trends. We used 0.4 mJ/cm$^2$ pump fluence. **c**, Equilibrium reflectivity at 90 K (black line) and reflectivity blue shifted by 0.1 THz (blue dashed line). **d**, Photoinduced reflectivity change $R_{pumped} - R_{eq}$ (red curve) along with reflectivity change due to blue shift of phonons (orange curve). The shaded regions under the curves are guide to the eye.

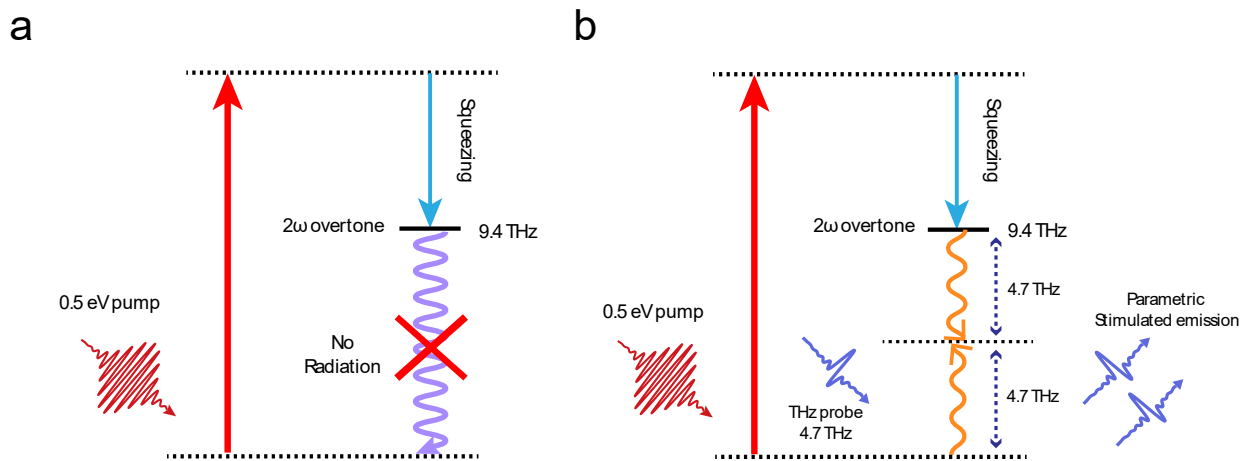

**Extended Data Fig. 8| THz emission possibilities.** Schematic of the squeezing process. **a**, The 9.4 THz $2\omega$ overtone is Raman active and thus cannot emit THz, resulting in no THz emission. **b**, On the contrary, it can emit stimulated radiation with the photon provided by the THz probe. As such, the THz probe is parametrically amplified.

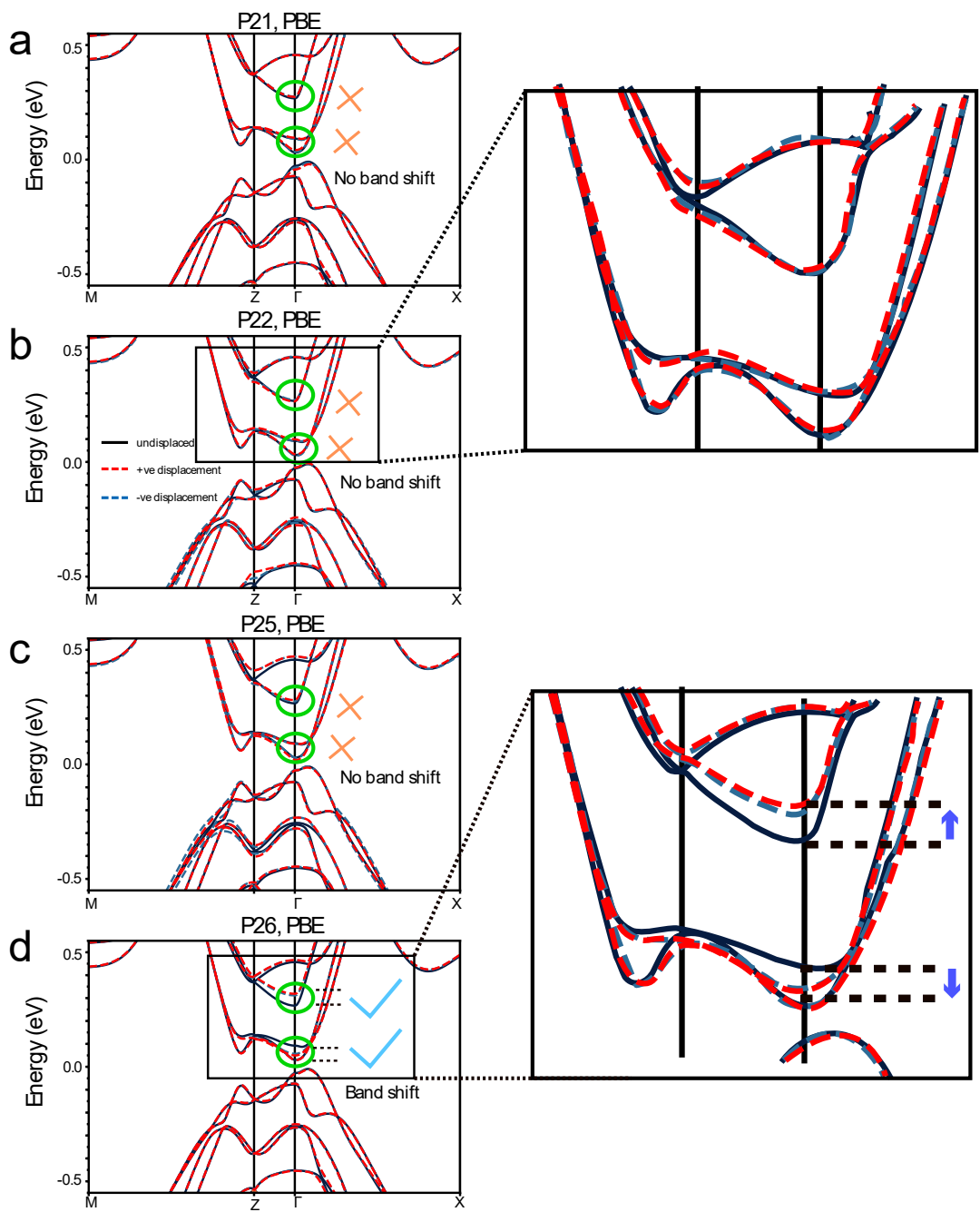

**Extended Data Fig. 9| DFT calculation.** Recalculated band structures in the monoclinic phase after displacement of the IR-phonon coordinate along the positive (red dashed line) and negative (dark blue dashed line) directions with respect to the equilibrium structure (black solid line) for **a,** mode 21, **b**, mode 22, **c,** mode 25 and **d**, mode 26 using the PBE functional. Green ellipses represent the shift in bands. Phonon mode 26 shows the largest renormalization and thus the strongest coupling to the band structure upon displacement along its eigenmode. Insets show the zoomed in profiles of the renormalized band structures. Blue arrows signify the direction of the energy shift.

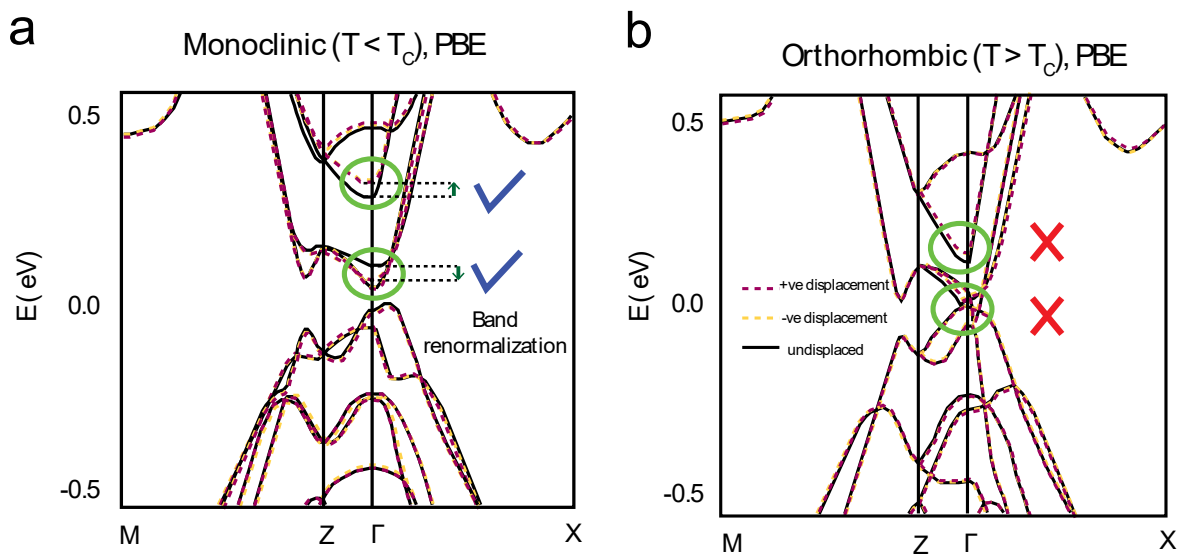

**Extended Data Fig. 10| Evolution of the 4.69 THz mode.** The temperature-dependence of the electron-phonon coupling for 4.69 THz $B_u$ mode and (mode 26) in the **a**, monoclinic and **b**, orthorhombic phases using the PBE functional. For the monoclinic phase, the electron-phonon coupling is strong and thus the shift in the bands owing to the displacement of the phonon coordinate along its eigenmode is large. Contrary to this, the band shift almost disappears for the orthorhombic phase. This demonstrates that the electron-phonon coupling is mediated by the low-temperature order parameter.

| Phonon | Symmetry | Frequency (THz) |
|:---:|:---:|:---:|
| 21 | $B_u$ | 4.24 |
| 22 | $A_u$ | 4.31 |
| 25 | $A_u$ | 4.63 |
| 26 | $B_u$ | 4.69 |

**Table 1| IR-active phonon modes near 4.5 THz.** Here there are four phonons with frequency close to 4.5 THz, frequency regime of our interest. Only the 4.69 THz $B_u$ mode is observed to be strongly coupled to the band structure.

**References and Notes**

41. Xie, X., Dai, J. & Zhang, X. C., Coherent control of THz wave generation in ambient air. *Phys. Rev. Lett.* **96**, 075005 (2006). doi:10.1103/PhysRevLett.96.075005

42. Dai, J., Karpowicz, N. & Zhang, X. C., Coherent polarization control of terahertz waves generated from two-color laser-induced gas plasma. *Phys. Rev. Lett.* **103**, 023001 (2009). doi:10.1103/PhysRevLett.103.023001

43. Werley, C. A., Teo, S. M. & Nelson, K. A., Pulsed laser noise analysis and pump-probe signal detection with a data acquisition car, *Rev. Sci. Instrum*. **82**, 123108 (2011). doi:10.1063/1.3669783

44. Kindt, J. T. & Schmuttenmaer, C. A., Theory of determination of the low-frequency time-dependent response function in liquids using time-resolved terahertz pulse spectroscopy. *J. Chem. Phys*. **110**, 8589-8596 (1999). doi:10.1063/1.478766

45. Coslovich, G., Kemper, A. F., Behl, S., Huber, B., Bechtel, H. A., Sasagawa, T. Martin, M. C., Lanzara, A. & Kaindl, R. A., Ultrafat dynamics of a vibrational symmetry breaking in a charge-ordered nickelate. *Sci. Adv.* **3**, e1600735 (2017). doi:10.1126/sciadv.1600735

46. Rajasekaran, S., Casandruc, E., Laplace, Y. Nicoletti, D., Gu, G.D., Clark, S. R., Jaksch, D. & Cavalleri, A., Parametric amplification of a superconducting plasma wave. *Nat. Phys.* **12**, 1012-1016 (2016). doi:10.1038/nphys3819

47. Kennes, D. M., Wilner, E. Y., Reichman, D. R. & Millis, A. J., Transient superconductivity from electronic squeezing of optically pumped phonons. *Nat. Phys*. **13**, 479-483 (2017). doi:10.1038/nphys4024

48. Garrett, G. A., Rojo, A. G., Sood, A. K., Whitaker, J. F. & Merlin, R., Vacuum squeezing of solids: Macroscopic quantum states driven by light pulses. *Science* **275**, 1638-1640 (1997). doi:10.1126/science275.5306.1638

49. Buzzi, M., Jotzu, G. Cavalleri, A., Cirac, J. I., Demler, E. A., Halperin, B. I., Lukin, M. D., Shi, T., Wang, Y. & Podolsky, D., Higgs-mediated optical amplification in a nonequilibrium superconductor. *Phys. Rev. X* **11**, 011055 (2021). doi:10.1103/PhysRevX.11.011055

50. Kresse, G. & Furthmüller, J., Efficient iterative schemes for *ab inito* total-energy calculations using a plane-wave basis set. *Phys. Rev. B* **54**, 11169-11186 (1996). doi:10.1103/PhysRevB.54.11169

51. Kresse, G. & Furthmüller, J., Efficiency of *ab inito* total-energy calculations for metals and semiconductors using a plane-wave basis set. *Comp. Mater. Sci.* **6 (1**), 15-50 (1996). doi:10.1016/0927-0256(96)00008-0

52. Kresse, G. & Hafner, J., *Ab initio* molecular dynamics for liquid metals. *Phys. Rev. B* **47**, 558-561 (1993). doi:10.1103/PhysRevB.47.558

53. Kresse, G. & Hafner, J., Norm-conserving and ultrasoft pseudopotentials for first-row and transition elements. *J. Phys.: Condens. Matter* **6**, 8245-8257 (1994). doi:10.1088/0953-8984/6/40/015

54. Kilmeš. J., Bowler, D. R. & Michaelides, A., Van der waals density functionals applied to solids. *Phys. Rev. B*. **83**, 195131 (2011). doi:10.1103/PhysRevB.83.195131

55. Kilmeš. J., Bowler, D. R. & Michaelides, A., Chemical accuracy for the van der Waals density functional. *J. Phys.: Condens. Matter* **22**, 022201 (2009). doi:10.1088/0953-8984/22/2/022201

56. Skelton, J. M., Burton, L. A., Jackson, A. J., Oba, F., Parker, S. C. & Walsh, A., Lattice dynamics of the tin sulphides $SnS_2$, SnS and $Sn_2S_3$: vibrational spectra and thermal transport. *Phys. Chem. Chem. Phys*. **19**, 12452-12465 (2017). doi:10.1039/C7CP01680H

57. Togo, A. & Tanaka, I., First principles phonon calculations in materials science. *Scr. Mater*. **108**, 1-5 (2015). doi:10.1016/j.scriptamat.2015.07.021